\documentclass{iopart}
\usepackage{iopams}
\usepackage{setstack}
\usepackage{graphicx,color,array,subfigure}
\usepackage{hyperref}
\hypersetup{
    bookmarks=true,         
    unicode=false,          
    pdftoolbar=true,        
    pdfmenubar=true,        
    pdffitwindow=false,     
    pdfstartview={FitH},    
    pdftitle={Shock waves in dispersive Eulerian fluids},    
    pdfauthor={Mark A. Hoefer},     
    pdfsubject={},   
    pdfcreator={Mark A. Hoefer},   
    pdfproducer={}, 
    pdfkeywords={dispersive shock wave} {Euler equations}, 
    pdfnewwindow=true,      
    colorlinks=false,       
    linkcolor=red,          
    citecolor=green,        
    filecolor=magenta,      
    urlcolor=cyan           
}

\definecolor{grn}{rgb}{0,0.5,0}
\eqnobysec 

\newcommand{\pd}[2]{\frac{\partial #1}{\partial #2}}

\newcommand{\eps}{\varepsilon}
\newcommand{\R}{\mathbb{R}}

\newcommand{\rhobar}{\overline{\rho}}
\newcommand{\ubar}{\overline{u}}
\newcommand{\rhomin}{\rho_\mathrm{min}}

\newcommand{\rhom}{\rho_\mathrm{m}}
\newcommand{\um}{u_\mathrm{m}}

\newcommand{\Deltav}{\Delta_\mathrm{v}}
\newcommand{\Deltas}{\Delta_\mathrm{s}}

\newcommand{\sigmacr}{\sigma_\mathrm{cr}}
\newcommand{\kcr}{k_\mathrm{cr}}
\newcommand{\tbr}{t_\mathrm{br}}
\newcommand{\kt}{\tilde{k}}
\newcommand{\alphat}{\tilde{\alpha}}
\newcommand{\omegat}{\tilde{\omega}}

\newtheorem{remark}{Remark}

\newcommand{\qed}{\nobreak \ifvmode \relax \else
      \ifdim\lastskip<1.5em \hskip-\lastskip
      \hskip1.5em plus0em minus0.5em \fi \nobreak
      \vrule height0.75em width0.5em depth0.25em\fi}

\begin{document}

\title{Shock waves in dispersive Eulerian fluids}
\author{M. A. Hoefer}
\address{Department of Mathematics, North Carolina State University,
    Raleigh, NC 27695}
\ead{mahoefer@ncsu.edu}

\begin{abstract}
  The long time behavior of an initial step resulting in a dispersive
  shock wave (DSW) for the one-dimensional isentropic Euler equations
  regularized by generic, third order dispersion is considered by use
  of Whitham averaging.  Under modest assumptions, the jump conditions
  (DSW locus and speeds) for admissible, weak DSWs are characterized
  and found to depend only upon the sign of dispersion (convex or
  concave) and a general pressure law.  Two mechanisms leading to the
  breakdown of this simple wave DSW theory for sufficiently large
  jumps are identified: a change in the sign of dispersion, leading to
  gradient catastrophe in the modulation equations, and the loss of
  genuine nonlinearity in the modulation equations.  Large amplitude
  DSWs are constructed for several particular dispersive fluids with
  differing pressure laws modeled by the generalized nonlinear
  Schr\"{o}dinger equation.  These include superfluids (Bose-Einstein
  condensates and ultracold Fermions) and ``optical fluids''.
  Estimates of breaking times for smooth initial data and the long
  time behavior of the shock tube problem are presented.  Numerical
  simulations compare favorably with the asymptotic results in the
  weak to moderate amplitude regimes.  Deviations in the large
  amplitude regime are identified with breakdown of the simple wave
  DSW theory.
\end{abstract}


\maketitle

%
%

\section{Introduction}
\label{sec:introduction}

\begin{figure}
  \centering
  \includegraphics[scale=1]{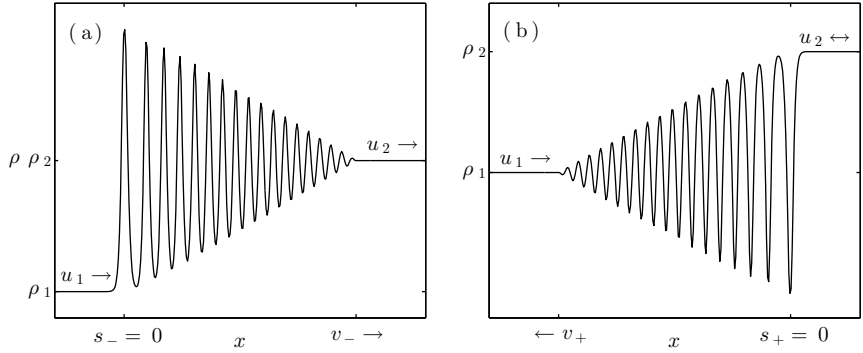}
  \caption{Density for the negative dispersion, $1^-$-DSW case (a) and
    positive dispersion, $1^+$-DSW case (b) with stationary soliton
    edge $s_+ = s_- = 0$ (see section \ref{sec:admiss-crit}).  The
    background flow velocities $u_1$, $u_2$ and linear wave edge
    velocities $v_+$, $v_-$ are also pictured.  In (a), backflow ($v_-
    < 0$) occurs while in (b), it is possible for the downstream flow
    to be negative when a vacuum point appears (see section
    \ref{sec:applications}).}
  \label{fig:pos_neg_dispersion}
\end{figure}

Nonlinear wave propagation in dispersive media with negligible
dissipation can lead to the formation of dispersive shock waves
(DSWs).  In contrast to classical, viscous shock waves which are
localized, rapid jumps in the fluid's thermodynamic variables, DSWs
exhibit an expanding oscillatory region
connecting two disparate fluid states.  A schematic depicting typical
left-going DSWs for positive and negative dispersion fluids is shown
in figure \ref{fig:pos_neg_dispersion}.  These structures are of
particular, current interest due to their recent observation in
superfluidic Bose-Einstein condensates (BECs) of cold atomic gases
\cite{dutton_observation_2001,simula_observations_2005,hoefer_dispersive_2006-1,chang_formation_2008,meppelink_observation_2009}
and nonlinear photonics
\cite{wan_dispersive_2007,jia_dispersive_2007,barsi_dispersive_2007,ghofraniha_shocks_2007,conti_observation_2009,wan_wave_2010,conforti_dispersive_2012,ghofraniha_measurement_2012}.
Dispersive shock waves also occur in a number of other dispersive
hydrodynamic type systems including water waves
\cite{chanson_current_2009} (known as undular hydraulic jumps or
bores), two-temperature collisionless plasma
\cite{taylor_observation_1970} (called collisionless shock waves), and
fluid interfaces in the atmosphere
\cite{smith_waves_1988,christie_morning_1992,rottman_atmospheric_2001}
and ocean \cite{holloway_internal_2001}.

The Whitham averaging technique
\cite{whitham_non-linear_1965,whitham_linear_1974} is a principle
analytical tool for the dispersive regularization of singularity
formation in hyperbolic systems; see e.g., the review
\cite{hoefer_dispersive_2009}.  The method is used to describe
  slow modulations of a nonlinear, periodic traveling wave.  Given an
  $n^{\mathrm{th}}$ order nonlinear evolution equation, implementation
  of the method requires the existence of a $n$-parameter family of
  periodic traveling wave solutions $\phi(\theta;\mathbf{p})$,
  $\mathbf{p}\in \R^n$ with period $L$ and phase $\theta$.
  Additionally, the evolution equation must admit $n-1$ conserved
  densities $\mathcal{P}_i[\phi]$ and fluxes $\mathcal{Q}_i[\phi]$,
  $i=1,2,\ldots,n-1$ corresponding to the conservation laws
\begin{equation}
  \label{eq:6}
  \frac{\partial}{\partial t}\mathcal{P}_i +
  \frac{\partial}{\partial x} \mathcal{Q}_i = 0 , \qquad i = 1,
  \ldots, n-1 .
\end{equation}
Assuming slow spatio-temporal evolution of the wave's parameters
$\mathbf{p}$, the conservation laws are then averaged over a period resulting
in the modulation equations
\begin{equation}
  \fl
  \label{eq:4}
  \left ( \frac{1}{L} \int_0^L
    \mathcal{P}_i[\phi(\theta;\mathbf{p})] \rmd\theta \right )_t 
  + \left ( \frac{1}{L} \int_0^L
    \mathcal{Q}_i[\phi(\theta;\mathbf{p})] \rmd\theta \right 
  )_x = 0 , \qquad i=1, \ldots, n-1.
\end{equation}
The $n$ Whitham modulation equations are completed by the addition
of the conservation of waves to (\ref{eq:4})
\begin{equation}
  \label{eq:5}
  k_t + \omega_x = 0, \qquad k = 2\pi/L = \theta_x,   c = - \theta_t ,
\end{equation}
a consistency condition ($\theta_{xt} = \theta_{tx}$) for the
application of modulation theory.  The Whitham equations are a set of
first order, quasi-linear partial differential equations (PDEs)
describing the slow evolution of the traveling wave's parameters
$\mathbf{p}$.

As laid out originally by Gurevich and Pitaevskii
\cite{gurevich_nonstationary_1974}, a DSW can be described by the
evolution of a free boundary value problem.  The boundary separates
the oscillatory, one-phase region, described by the Whitham equations,
from non-oscillatory, zero-phase regions, described by the
dispersionless evolution equation.  The regions are matched at phase
boundaries by requiring that the average of the one-phase solution
equals the zero-phase solution.  Thus, the free boundary is determined
along with the solution.  There are two ways for a one-phase wave to
limit to a zero-phase solution.  In the vicinity of the free boundary,
either the oscillation amplitude goes to zero (harmonic limit) or the
oscillation period goes to infinity, corresponding to a localization
of the traveling wave (soliton limit).  The determination of which
limiting case to choose at a particular phase boundary requires
appropriate admissibility criteria, analogous to entropy conditions
for classical shock waves.

Riemann problems consisting of step initial data are an analytically
tractable and physically important class to study.  For a system of
two genuinely nonlinear, strictly hyperbolic conservation laws, the
general solution of the Riemann problem consists of three constant
states connected by two self-similar waves, either a rarefaction or a
shock \cite{lax_hyperbolic_1973,smoller_shock_1994}.  This behavior
generalizes to dispersive hydrodynamics so, borrowing terminology from
classical shock theory, it is natural to label a left(right)-going
wave as a 1(2)-DSW or 1(2)-rarefaction.  See figure
\ref{fig:pos_neg_dispersion} for examples of $1^\pm$-DSWs where the
$sign$ corresponds to positive or negative dispersion.  For a DSW
resulting from the long time evolution of step initial conditions, the
oscillatory boundaries are straight lines.  These leading and trailing
edge speeds can be determined in terms of the left and right constant
states, analogous to the Rankine-Hugoniot jump conditions of classical
gas dynamics.  Whitham modulation theory for DSWs was initially
developed for integrable wave equations.  Integrability in the context
of the modulation equations \cite{tsarev_poisson_1985} implies the
existence of a diagonalizing transformation to Riemann invariants
where the Riemann problem for the hyperbolic modulation equations
could be solved explicitly for a self-similar, simple wave
\cite{gurevich_nonstationary_1974,gurevich_dissipationless_1987}.  The
two DSW speeds at the phase boundaries coinciding with the soliton and
harmonic limits are the characteristic speeds of the edges of the
simple wave.  Thus, the dispersive regularization of breaking in a
hydrodynamic system is implemented by the introduction of additional
conservation laws (the Whitham equations) that admit a global
solution.  An important innovation was developed by El
\cite{el_resolution_2005} whereby the DSW's trailing and leading edge
speeds could be determined without solving the full set of modulation
equations.  The Whitham-El DSW construction relies on the existence of
a simple wave solution to the full, strictly hyperbolic and genuinely
nonlinear modulation equations, but does not require its complete
determination, hence analytical results are available even for
non-integrable equations.

In this work, the one-dimensional (1D) isentropic Euler equations are
regularized by a class of third order dispersive terms, modeling
several of the aforementioned physical systems.  The time to breaking
(gradient catastrophe) for smooth initial data is numerically found to
fall within bounds predicted by the dispersionless Euler equations.
In order to investigate dynamics post-breaking, the long time
resolution of the Riemann problem is considered.  The DSW locus
relating upstream, downstream flow configurations and the DSW speeds
for admissible weak shocks are determined explicitly for generic,
third order dispersive perturbations.  The results depend only upon
the sign of the dispersion ($\mathrm{sgn}\, \omega_{kk}$) and the
general pressure law assumed.  A fundamental assumption in the DSW
construction is the existence of an integral curve (simple wave) of
the Whitham modulation equations connecting the upstream and
downstream states in an averaged sense.  Explicit, verifiable criteria
for the breakdown of the simple wave assumption are given.  The
regularization for large amplitude DSWs depends upon the particular
form of the dispersion.  Thus, DSWs are explicitly constructed for
particular pressure laws and dispersive terms of physical origin
including a generalized Nonlinear Schr\"{o}dinger (gNLS) equation
modeling superfluids and nonlinear optics.  Comparisons with a
dissipative regularization are presented in order to highlight the
differences between viscous and dispersive shock waves.

The outline of this work is as follows.  Section
\ref{sec:disp-euler-equat} presents the general dispersive Euler model
and assumptions to be considered, followed by section
\ref{sec:example-systems} outlining the gNLS and other dispersive
Eulerian fluid models.  Background sections
\ref{sec:dispersionless-limit} and \ref{sec:normal-dsws} review the
theory of the hyperbolic, dispersionless system and the Whitham-El
method of DSW construction, respectively.  A detailed analysis of DSW
admissibility criteria and the breakdown of the simple wave assumption
are undertaken in section \ref{sec:admiss-crit} followed by the
complete characterization of admissible weak DSWs in section
\ref{sec:weak-dsws}.  The theory of the breaking time for the gNLS
model is shown to agree with numerical computation in section
\ref{sec:dsw-breaking-time}.  Large amplitude DSWs for the gNLS
equation are studied in section \ref{sec:applications}.  The
manuscript is completed by conclusions and an appendix on the
numerical methods utilized.

\section{Dispersive Euler Equations and Assumptions}
\label{sec:disp-euler-equat}

The 1D dispersive Euler equations considered in this work are, in
non-dimensional form
\begin{eqnarray}
  \label{eq:1}
  \eqalign{
  \rho_t + (\rho u)_x = 0, \\
  (\rho u)_t + \left [ \rho u^2 + P(\rho) \right ]_x =
  [D(\rho,u)]_x , \qquad -\infty < x < \infty,}
\end{eqnarray}
where $\rho$ is a fluid density, $u$ is the velocity, and $D_x$ is the
(conservative) dispersive term.  Formally setting $D = 0$ gives the
hydrodynamic approximation, valid until gradient catastrophe when the
dispersion acts to regularize the singular behavior.  The
dispersionless, hyperbolic, isentropic Euler equations are known as
the $P$-system whose weak solutions to the Riemann problem are
well-known \cite{wagner_equivalence_1987,smoller_shock_1994}.  Here,
the long time behavior of the dispersively regularized Riemann problem
is analyzed.  By the formal rescaling $X = \eps x$, $T = \eps t$, 
\begin{eqnarray*}
  \eqalign{
    \rho_T + (\rho u)_X = 0, \\
    (\rho u)_T + \left [ \rho u^2 + P(\rho) \right ]_X =
    \eps^2 [D(\rho,u)]_X , \qquad -\infty < X < \infty,
  },
\end{eqnarray*}
the long time ($t \gg 1$) behavior of the dispersive Euler equations
in the independent variables $(X,T)$ is recast as a small dispersion
($\eps^2 \ll 1$) problem.  Due to the oscillatory nature of the small
dispersion limit, it is necessarily a weak limit as shown rigorously
by Lax, Levermore, and Venakides for the Korteweg-deVries equation
(KdV)
\cite{lax_small_1983-2,lax_small_1983,lax_small_1983-1,venakides_zero-dispersion_1985}.
In this work, the multiscale Whitham averaging technique will be used
to study the behavior of the dispersive Euler equations (\ref{eq:1})
for sufficiently large time and long waves.

The Whitham-El DSW simple wave closure method
\cite{el_resolution_2005} is used to construct DSWs under the
following assumptions.
\begin{itemize}
\item[\textbf{A1}] \textbf{(sound speed)} The pressure law $P =
  P(\rho)$ is a smooth, monotonically increasing function of $\rho$,
  $P'(\rho) > 0$ for $\rho > 0$ so that the speed of sound
  \begin{equation*}
    c_0 = c(\rho_0) \equiv \sqrt{P' (\rho_0)},
  \end{equation*}
  is real and the local Mach number
  \begin{equation*}
    M_0 \equiv \frac{| u_0 |}{c(\rho_0)},
  \end{equation*}
  is well-defined.  It will also be assumed that the pressure is
  convex $P''(\rho) > 0$ for $\rho > 0$ so that $c'(\rho) > 0$.
\item[\textbf{A2}] \textbf{(symmetries)}
  Equations~(\ref{eq:1}) admit the Galilean invariance:
  \begin{equation*}
    D(\rho,u-u_0)(x-u_0 t,t) =
    D(\rho,u)(x,t) ,
  \end{equation*}
  for all $u_0 \in \R$ and exhibit the sign inversion
  \begin{equation}
    \label{eq:67}
    D(\rho,-u)(-x,t) = -D(\rho,u)(x,t),
  \end{equation}
  so that (\ref{eq:1}) are invariant with
  respect to $x \to -x$, $u \to -u$.  
\item[\textbf{A3}] \textbf{(dispersive operator)} The dispersive term
  $(D[\rho,u])_x$ is a differential operator with $D$ of second order
  in spatial and/or mixed partial derivatives such that the system
  (\ref{eq:67}) has the real-valued dispersion relation
  \begin{equation}
    \label{eq:15}
    \omega = u_0 k \pm \omega_0(k,\rho_0) ,
  \end{equation}
  with two branches found by linearizing about the uniform background
  state $\rho = \rho_0$, $u = u_0$ with small amplitude waves
  proportional to $\exp[\rmi(k x - \omega t)]$.  The appropriate
  branch of the dispersion relation is fixed by the $\pm$ in
  (\ref{eq:15}) with $\omega_0(k,\rho_0) \ge 0$ for $k \ge 0$, $\rho_0
  \ge 0$.  The dispersion relation has the long wave expansion
  \begin{equation}
    \label{eq:80}
    \omega_0(k,\rho_0) = c_0 k + \mu k^3 + o(k^3), \qquad k
    \to 0, \qquad \mu \ne 0 .
  \end{equation}
  The sign of the dispersion is $\mathrm{sgn} ~ \omega_0''(k;\rho_0)$
  for $k > 0$.  Using (\ref{eq:80}) and the convexity or concavity of
  $\omega_0$ as a function of $k$, one finds
  \begin{equation*}
    \mathrm{sgn} \left ( \frac{\omega_0(k,\rho_0)}{k} \right )_k =
    \mathrm{sgn} \frac{\partial^2 \omega_0}{\partial k^2} (k,\rho_0) .
  \end{equation*}
  Therefore, positive dispersion corresponds to increasing phase and
  group velocities with increasing $k$ while negative dispersion leads
  to decreasing phase and group velocities.
\item[\textbf{A4}] \textbf{(Whitham averaging)} Equations (\ref{eq:1})
  are amenable to Whitham averaging whereby a DSW
  can be described by a slowly varying, single-phase traveling wave.
  This requires
  \begin{enumerate}
  \item The system possesses at least three conservation laws.  The
    mass and momentum equations in (\ref{eq:1}) account
    for two.  An additional conserved quantity is required.
  \item There exists a four parameter family of periodic traveling
    waves parametrized by, for example, the wave amplitude $a$, the
    wavenumber $k$, the average density $\rhobar$, and the average
    velocity $\ubar$ limiting to a trigonometric wave for small
    amplitude and a solitary wave for small wavenumber.  In the cases
    considered here, the periodic traveling wave manifests as a
    solution of the ordinary differential equation (ODE) $(\rho')^2 =
    G(\rho)$ where $G$ is smooth as it varies over three simple, real
    roots.  Two roots coincide in the small amplitude and solitary
    wave limits.
  \end{enumerate}
\item[\textbf{A5}] \textbf{(Simple wave)} The Whitham-El method
  requires the existence of a self-similar simple wave solution to the
  four Whitham modulation equations (the averaged conservation laws
  and the conservation of waves).  For this, the modulation equations
  must be strictly hyperbolic and genuinely nonlinear.
\end{itemize}

Assumption A1 provides for a modulationally stable, hydrodynamic long
wave limit.  The symmetry assumptions in A2 are for convenience and
could be neglected.  As will be demonstrated in section
\ref{sec:example-systems}, A3 is a reasonable restriction still
allowing for a number of physically relevant dispersive fluid models.
The assumptions in A4 and A5 allow for the application of the
Whitham-El method.  While the assumptions in A4 are usually
verifiable, A5 is often assumed.  Causes of the breakdown of
assumptions A3 (unique dispersion sign) and A5 (genuine nonlinearity)
are identified and associated with extrema in the DSW speeds as either
the left or right density is varied.

The nonstationary DSW considered here is the long time resolution of
an initial jump in the fluid density and velocity, the Riemann problem
\begin{equation}
  \label{eq:77}
  u(x,0) = \left \{
    \begin{array}{ll}
      u_1 & x < 0 \\
      u_2 & x > 0
    \end{array} \right . , \qquad 
  \rho(x,0) = \left \{
    \begin{array}{ll}
      \rho_1 & x < 0 \\
      \rho_2 & x > 0
    \end{array} \right . ,
\end{equation}
where $u_j \in \R$, $\rho_j \ge 0$.

\section{Example Dispersive Fluids}
\label{sec:example-systems}

The dispersive Euler equations (\ref{eq:67}) model a number of
dispersive fluids including, among others, superfluids and optical
fluids.  The particular model equations described below were chosen
because they incorporate different pressure laws and allow for
different signs of the dispersion, key distinguishing features of
Eulerian dispersive fluids and their weak dispersion regularization.

\subsection{gNLS Equation}
\label{sec:gnls}

The generalized, defocusing nonlinear Schr\"{o}dinger equation
\begin{eqnarray}
  \label{eq:83}
  \eqalign{
  \rmi \psi_t = -\frac{1}{2} \psi_{xx} +
  f(|\psi|^2) \psi, \\
  f(0) = 0, \qquad f(\rho) > 0, \qquad \rho > 0,}
\end{eqnarray}
or gNLS, describes a number of physical systems.  For example, the
``polytropic superfluid''
\begin{equation}
  \label{eq:164}
  f(\rho) = \rho^p, \qquad p > 0,
\end{equation}
corresponds to the cubic NLS when $p=1$ that describes a repulsive BEC
and intense laser propagation through optically defocusing (normal
dispersion) media.  The model (\ref{eq:164}) with $p = 2/3$ describes
a zero temperature Fermi gas near unitarity
\cite{giorgini_theory_2008,csordas_gradient_2010} which is of special
significance as recent experiments have been successfully interpreted
with both dissipative \cite{joseph_observation_2011} and dispersive
\cite{salasnich_dynamical_2012} regularizations.  Moreover, the regime
$2/3 < p < 1$ describes the so-called BEC-BCS transition in ultracold
Fermi gases \cite{ketterle_making_2008}.  The quintic NLS case, $p =
2$, models three-body interactions in a BEC
\cite{kevrekidis_emergent_2008,chen_quintic_2011}.  
A BEC confined to a cigar shaped trap exhibits effective 1D behavior
that is well-described by the non-polynomial nonlinearity
\cite{salasnich_effective_2002,mateo_effective_2008}
\begin{equation}
  \label{eq:165}
  f(\rho) = \frac{2\sqrt{1 + \gamma \rho} - 2}{\gamma}, \qquad \gamma >
  0 ,
\end{equation}
here scaled so that $f(\rho) \to \rho$, $\gamma \to 0^+$.  In spatial
nonlinear optics, photorefractive media corresponding to
\cite{segev_spatial_1992,christodoulides_bright_1995}
\begin{equation}
  \label{eq:84}
  f(\rho) = \frac{\rho}{1 + \gamma \rho}, \qquad \gamma > 0,
\end{equation}
is of particular interest due to recent experiments exhibiting DSWs
\cite{wan_dispersive_2007,barsi_dispersive_2007,jia_dispersive_2007,barsi_spatially_2012,jia_rayleightaylor_2012}.
For $0 < \gamma \ll 1$, the leading order behavior of (\ref{eq:165})
and (\ref{eq:84}) correspond to the cubic NLS.

The complex wavefunction $\psi$ can be interpreted in the dispersive fluid
context by use of the Madelung transformation
\cite{madelung_quantentheorie_1927}
\begin{equation}
  \label{eq:85}
  \psi = \sqrt{\rho} \rme^{\rmi \phi}, \qquad u = \phi_x .
\end{equation}
Using (\ref{eq:85}) in (\ref{eq:83}) and equating real and imaginary
parts results in the dispersive Euler equations (\ref{eq:1}) with
\begin{eqnarray}
  \label{eq:86}
  \eqalign{
    P(\rho) = \int_0^\rho \tilde{\rho} f'(\tilde{\rho}) \rmd
    \tilde{\rho} , \qquad
    c(\rho) = \sqrt{\rho f'(\rho)} ,\\
    [D(\rho,u)]_x = \frac{1}{4} \left [
      \rho \left (   
        \ln \rho \right )_{xx} \right  ]_x = \frac{\rho}{2} \left [
      \frac{(\sqrt{\rho})_{xx}}{\sqrt{\rho}}\right ]_x .}
\end{eqnarray}
The dispersive regularization of (\ref{eq:1})
corresponds to the semi-classical limit of (\ref{eq:83}), which, in
dimensional units corresponds to $\hbar \to 0$ for quantum many body
systems.  In applications, the dispersive regularization coincides
with a strongly interacting BEC or a large input optical intensity.

Assumption A1 restricts the admissible nonlinearity $f$ to those
satisfying
\begin{equation}
  \label{eq:139}
  f'(\rho) > 0, \qquad \left ( \rho f'(\rho) \right )' > 0, \qquad
  \rho > 0, 
\end{equation}
which is realized by (\ref{eq:164}), (\ref{eq:165}) generally and for
(\ref{eq:84}) when $\gamma \rho < 1$.  Assumptions in A2 are well-known
properties of the gNLS equation \cite{sulem_nonlinear_1999}.
Assumption A3 is clear from (\ref{eq:86}) and the dispersion relation
is
\begin{equation}
  \label{eq:89}
  \omega_0(k,\rho) = k  \sqrt{c^2 + k^2/4} \sim c k + \frac{1}{8 c}
  k^3, \qquad |k|  \ll c .
\end{equation}
The dispersion is positive because $\omega_{0_{kk}}(k;\rho) > 0$ for $k >
0$, $\rho \ge 0$.

Inserting the traveling wave ansatz
\begin{equation*}
  \rho = \rho(x - Vt), \qquad u = u(x-Vt),
\end{equation*}
into (\ref{eq:1}) with (\ref{eq:86}) and integrating
twice leads to
\numparts
\begin{eqnarray}
  \label{eq:90}
  u = V + \frac{A}{\rho}, \\
  \label{eq:91}
  \left ( \rho' \right )^2 = 8 \left [ \rho \int_{\rho_1}^\rho
    f(\tilde{\rho}) \rmd \tilde{\rho} + B \rho^2 + C \rho -
    \frac{A^2}{2} \right ] 
  \equiv G(\rho) .
\end{eqnarray}
\endnumparts
It is assumed that $G$ has three real roots $\rho_1 \le \rho_2 \le
\rho_3$ related to the integration constants $A$, $B$, and $C$ so
that, according to a phase plane analysis, a periodic wave exists with
maximum and minimum densities $\rho_2$ and $\rho_1$, respectively.
The fourth arbitrary constant, due to Galilean invariance, is the wave
speed $V$.  In addition to mass and momentum conservation, an
additional energy conservation law exists
\cite{jin_semiclassical_1999} which reads
\begin{eqnarray*}
  \eqalign{
    \mathcal{E} \equiv \frac{\rho u^2}{2} + \frac{\rho_x^2}{8 \rho} +
    \int_0^\rho f(\tilde{\rho}) \rmd \tilde{\rho} , \\
    \mathcal{E}_t + \left \{ u [ \mathcal{E} + P(\rho) ]
    \right \}_x = \frac{1}{4}  \left [ u \rho_{xx} -
      \frac{(\rho u)_x \rho_x}{\rho} \right ]_x ,}
\end{eqnarray*}
hence the assumptions in A4 are satisfied.  The hyperbolicity of the
Whitham equations can only be determined by their direct study.  The
genuine nonlinearity of the system will be discussed in section
\ref{sec:admiss-crit}.  It will be helpful to note the solitary wave
amplitude/speed relation which results from the boundary conditions
for a depression (dark) solitary wave
\begin{equation*}
  u_0 \equiv \lim_{|\xi| \to \infty} u(\xi), \qquad \rho_0 \equiv
  \lim_{|x| \to \infty} \rho(\xi), \qquad \rhomin \equiv \min_{\xi \in
    \R} \rho(\xi) .
\end{equation*}
A phase plane analysis of (\ref{eq:91}) implies that the roots of $G$
satisfy $\rho_1 = \rhomin$, $\rho_2 = \rho_3 = \rho_0$ resulting in
the solitary wave speed $s = V$ satisfying
\begin{equation}
  \label{eq:93}
  (s - u_0)^2 = \frac{2 \rhomin}{(\rho_0 - \rhomin)^2} \left [ (\rho_0 -
    \rhomin ) f(\rho_0) - \int_{\rhomin}^{\rho_0} f(\tilde{\rho}) \rmd
    \tilde{\rho}  \right ] .
\end{equation}
The soliton profile can be determined by integration of (\ref{eq:91}).

Dispersive shock waves for the gNLS equation have been studied for the
pure NLS case \cite{gurevich_nonlinear_1990,el_general_1995} as well
as in 1D photorefractive media \cite{el_theory_2007} and the
cubic-quintic case
\cite{crosta_bistability_2011,crosta_crossover_2012}.  A general DSW
analysis will be presented in section \ref{sec:applications}.

\subsection{Other Systems}
\label{sec:other-systems}

The gNLS equation exhibits positive dispersion.  Two additional examples
are briefly given here with negative dispersion.

\textbf{Two-temperature collisionless plasma:} The dynamics of the
ionic component of a two-temperature unmagnetized plasma
\cite{karpman_non-linear_1974} satisfy the dispersive Euler equations
with
\begin{eqnarray*}
  \eqalign{
  P(\rho) = \rho, \qquad
  c(\rho) = 1, \\
  D(\rho,u) = \frac{1}{2} \phi_x^2 - \phi_{xx} , \qquad
  -\phi_{xx}   =  \rho - \rme^\phi.}
\end{eqnarray*}
The electronic potential $\phi$ introduces nonlocal dispersion with
dispersion relation
\begin{equation*}
  \omega_0(k,\rho) = \frac{k}{\sqrt{1 + k^2/\rho}} \sim k - \frac{k^3}{2\rho_0},
  \qquad |k| \ll 1 .
\end{equation*}
It can be shown that $\omega_{kk} < 0$, $k > 0$ thus the system
exhibits negative dispersion.

This system has been analyzed in detail \cite{gurevich_nonlinear_1990}
and satisfies assumptions A1-A4.  Large amplitude dispersive shock
waves were constructed in \cite{el_resolution_2005} under the
assumptions of A5.

\textbf{Fully nonlinear shallow water:} Shallow waves in an ideal
fluid with no restriction on amplitude satisfy the generalized Serre
equations (also referred to as the Su-Gardner or Green-Naghdi
equations)
\cite{serre_contribution_1953,su_korteweg-vries_1969,green_derivation_1976,dias_fully-nonlinear_2010}
with
\begin{eqnarray}
  \label{eq:96}
  \eqalign{
  P(\rho) = \frac{1}{2} \rho^2, \qquad
  c(\rho) = \sqrt{\rho}, \\
  D(\rho,u) = \frac{1}{3} \left [ \rho^3 \left ( 
      u_{tx} + uu_{xx} -
      u_x^2 \right ) \right ]  + \sigma
  \left (
    \rho \rho_{xx} - 
    \frac{1}{2} \rho_x ^2 \right ) .}
\end{eqnarray}
The density $\rho$ corresponds to the free surface fluid height and
$u$ is the vertically averaged horizontal fluid velocity.  The Bond
number $\sigma \ge 0$ is proportional to the coefficient of surface
tension.  The dispersion relation is
\begin{eqnarray*}
  \eqalign{
  \omega_0(k,\rho) &= k \left (\rho\frac{1 + \sigma k^2}{1 + \rho^2
      k^2/3} \right )^{1/2} \\
  &\sim  \sqrt{\rho}  \left ( k +
    \frac{3\sigma - \rho^2}{6} k^3 \right 
  ), \qquad k \to 0 .}
\end{eqnarray*}
The sign of the dispersion changes when $\omega_{kk} = 0$
corresponding to the critical values 
\begin{equation*}
  \sigmacr = \frac{\rho^2}{3} \qquad \mathrm{or} \qquad  \kcr = \frac{1}{\rho}
  \left ( 3 + 3\sqrt{1+\rho^2/\sigma} \right )^{1/2} .
\end{equation*}
The critical value $\sigmacr$ expresses the fact that shallow water
waves with weak surface tension effects, $\sigma < \sigmacr$, exhibit
negative dispersion for sufficiently long waves ($k < \kcr$) and
support elevation solitary wave solutions.  Strong surface tension,
$\sigma > \sigmacr$, corresponds to positive dispersion and can yield
depression solitary waves.  Assumptions A1-A4 hold
\cite{dias_fully-nonlinear_2010}.  DSWs in the case of zero surface
tension $\sigma = 0$ were studied in \cite{el_unsteady_2006}.

The Serre equations (\ref{eq:96}) with $\sigma = 0$ and a model
  of liquid containing small gas bubbles
  \cite{wijngaarden_one-dimensional_1972} can be cast in Lagrangian
  form to fit into the framework of ``continua with memory''
  \cite{gavrilyuk_generalized_2001}.  The Whitham modulation equations
  for these dispersive Eulerian fluids were studied in
  \cite{gavrilyuk_large_1994}.  Explicit, sufficient conditions for
  hyperbolicity of the modulation equations were derived.

The properties of DSWs for these systems will be discussed in section
\ref{sec:large-amplitude-dsws}.

\section{Background:  Dispersionless Limit}
\label{sec:dispersionless-limit}

The analysis of DSWs for (\ref{eq:1}) requires an
understanding of the dispersionless limit
\begin{eqnarray}
  \label{eq:121}
  \eqalign{
  \rho_t +  (\rho u)_x = 0, \\
  (\rho u)_t + \left [ \rho u^2 
    + P(\rho) \right ]_x = 0 ,}
\end{eqnarray}
corresponding to $D \equiv 0$.  Equations (\ref{eq:121}) are the
equations of compressible, isentropic gas dynamics with pressure law
$P(\rho)$ \cite{liepmann_elements_1957}.  They are hyperbolic and
diagonalized by the Riemann invariants (see
e.g.~\cite{courant_supersonic_1948})
\begin{equation}
  \label{eq:17}
  r_1 = u - \int^\rho \frac{c(\rho')}{\rho'} \rmd \rho', \qquad r_2
  = u + \int^\rho \frac{c(\rho')}{\rho'} \rmd \rho' ,
\end{equation}
with the characteristic velocities
\begin{equation}
  \label{eq:120}
  \lambda_1 = u - c(\rho), \qquad
  \lambda_2 = u +
  c(\rho) ,
\end{equation}
so that
\begin{equation}
  \label{eq:196}
  \pd{r_j}{t} + \lambda_j \pd{r_j}{x} = 0 , \qquad j = 1, 2.
\end{equation}
By monotonicity of
\begin{equation*}
  g(\rho) = \int^\rho \frac{c(\rho')}{\rho'} \rmd \rho',
\end{equation*}
the inversion of (\ref{eq:17}) is achieved via
\begin{equation}
  \label{eq:314}
  u = \frac{1}{2} ( r_1 + r_2), \qquad \rho = g^{-1} \left (
    \frac{1}{2} ( r_2 - r_1) \right ) .
\end{equation}

In what follows, an overview of the properties of equations
(\ref{eq:121}) is provided for both the required
analysis of DSWs and for the comparison of classical and dispersive
shock waves.

\subsection{Breaking Time}
\label{sec:breaking-time}

Smooth initial data may develop a singularity in finite time.  The
existence of Riemann invariants (\ref{eq:17}) allows for estimates of
the breaking time at which this occurs.  In what follows, Lax's
breaking time estimates \cite{lax_development_1964} are applied to the
system (\ref{eq:196}) with smooth initial data.

Lax's general approach for $2 \times 2$ hyperbolic systems is to
reduce the Riemann invariant system (\ref{eq:196}) to the equation
\begin{eqnarray}
  \label{eq:303}
  z' = -a(t) z^2,  \qquad z(0) = m,
\end{eqnarray}
along a characteristic family, $' \equiv \pd{}{t} + \lambda_i
\pd{}{x}$, and then bound the breaking time by comparison with an
autonomous equation via estimates for $a$ and $m$ in terms of initial
data for $r_{1}$, $r_2$.

Following Lax \cite{lax_development_1964}, integration along the
1-characteristic family in (\ref{eq:303}) leads to $z =
\rme^h \partial r_1/\partial x$, $a = \rme^{-h} \partial
\lambda_1/ \partial r_1 $ and $h(r_1,r_2)$ satisfies
\begin{equation*}
  \pd{h}{r_{2}} = \frac{\pd{\lambda_{1}}{r_{2}}}{\lambda_1 - \lambda_2} .
\end{equation*}
By direct computation with eqs.~(\ref{eq:17}), (\ref{eq:120}), one can
verify the following
\numparts
\begin{eqnarray}
  \label{eq:312}
  h = \frac{1}{2} \ln \left [  \frac{c(\rho)}{\rho} \right ] ,\\
  \label{eq:135}
  a = \rme^{-h} \pd{\lambda_1}{r_1} = \frac{c(\rho) + \rho 
    c'(\rho)}{2 c(\rho)} \left [ \frac{\rho}{c(\rho)}
  \right ]^{1/2}   , \\
  \label{eq:298}
  z = \rme^{h} \pd{r_1}{x} = \pd{r_1}{x} \left [ \frac{c(\rho)}{\rho} \right ]^{1/2} .
\end{eqnarray}
\endnumparts
The initial data for $r_1$ and $r_2$ are assumed to be smooth and
bounded so that they satisfy
\begin{equation}
  \label{eq:205}
  \underline{r_1} \le r_1(x,t) \le \overline{r_1}, \qquad
  \underline{r_2} \le r_2(x,t) \le \overline{r_2} , \qquad 0 \le t <
  \tbr .
\end{equation}
Assuming $\rho > 0$ (non-vacuum conditions), then (\ref{eq:135})
implies $a > 0$ and $z$ is decreasing along the 1-characteristic.
Bounds for $a(t)$ are defined as follows
\begin{eqnarray}
  \label{eq:306}
  A = \min_{\rho \in R_A} a, \qquad
  B = \max_{\rho \in R_B} a,
\end{eqnarray}
where $R_A$ and $R_B$ are intervals related to the bounds on the
initial data, chosen shortly.  The initial condition $m$ is chosen as
negative as possible
\numparts
\begin{eqnarray}
  \label{eq:315}
  \fl
  x_0 = \arg \min_{x\in \R} z(0) = \left . \arg \min_{x\in \R}  \left [ \pd{u}{x} - \frac{c(\rho)}{\rho}
      \pd{\rho}{x} \right ] \left [
      \frac{c(\rho)}{\rho} \right 
    ]^{1/2} \right |_{t=0} ,\\
  \label{eq:316}
  \fl
  m = \min_{x \in \R} z(0) = \left .  \left [ \pd{u}{x} - \frac{c(\rho)}{\rho}
      \pd{\rho}{x} \right ] \left [ \frac{c(\rho)}{\rho} \right
    ]^{1/2} \right |_{(x,t) = (x_0,0)}.
\end{eqnarray}
\endnumparts
These estimates lead to the following bounds on the breaking time
$\tbr$
\begin{equation}
  \label{eq:319}
  -\frac{1}{m B} \le \tbr \le - \frac{1}{mA} .
\end{equation}

It is still necessary to provide the intervals $R_A$ and $R_B$.  The
possible values of $r_1$ and $r_2$ in (\ref{eq:205}) and the
monotonicity of the transformation for $\rho$ in (\ref{eq:314})
suggests taking the full range of possible values $R_A = R_B =
[g^{-1}((\underline{r_2} - \overline{r_1})/2), g^{-1}((\overline{r_2}
- \underline{r_1})/2)]$.  However, this choice does not provide the
sharpest estimates in (\ref{eq:319}).  The idea is to use the fact
that $r_1$ is constant along 1-characteristics.  The choice for $m$ in
(\ref{eq:316}) suggests taking $r_1 = r_1(x_0,0)$ and allowing $r_2$
to vary across its range of values.  While the optimal $m$ is
associated with this characteristic, it does not necessarily provide
the optimal estimates for $A$ or $B$.  A calculation shows
\begin{equation*}
  \pd{a}{r_1} = -\frac{1}{8 c^3} \left ( \frac{\rho}{c} \right )^{1/2}
  \left (c^2 - 4 \rho c c' - 2 \rho^2 c c'' + 3 \rho^2 {c'}^2 \right )
  .
\end{equation*}
It can be verified that $\partial a/\partial r_1 \le 0$ for the
example dispersive fluids considered here.  In this case, any
characteristic with $r_1 < r_1(x_0,0)$ can cause $a$ to
\emph{increase}, leading to a larger $A$ and a tighter bound in
(\ref{eq:319}).  If $r_1 > r_1(x_0,0)$, then $a$ may \emph{decrease},
leading to a smaller $B$ and a tighter bound on $\tbr$.  Combining
these deductions leads to the choices
\numparts
\begin{eqnarray}
  \label{eq:320}
  R_A = \left [ g^{-1}\left ( \frac{\underline{r_2} - r_1(x_0,0)}{2}
    \right ) , g^{-1}\left ( \frac{\overline{r_2}
        - \underline{r_1}}{2} \right ) \right ], \\
  \label{eq:321}
  R_B = \left [ g^{-1}\left ( \frac{\underline{r_2} - \overline{r_1}}{2}
    \right ) , g^{-1}\left ( \frac{\overline{r_2}
        - r_1(x_0,0)}{2} \right ) \right ] ,
\end{eqnarray}
\endnumparts
when $\partial a/ \partial r_1 < 0$.

In summary, given initial data satisfying (\ref{eq:205}), the point
$x_0$ and $m$ are determined from (\ref{eq:315}) and (\ref{eq:316}).
If $m > 0$, then there is no breaking.  Otherwise, after verifying
$\partial a/ \partial r_1 < 0$, the sets $R_A$ and $R_B$ are defined
via (\ref{eq:320}) and (\ref{eq:321}) leading to $A$ and $B$ in
(\ref{eq:306}).  The breaking time bounds are given by (\ref{eq:319}).
A similar argument integrating along the 2-characteristic field yields
another estimate for the breaking time $\tbr$.  The only changes are
in (\ref{eq:315}) and (\ref{eq:316}) where the minus sign goes to a
plus sign and the choices for $R_A$ and $R_B$ reflect $r_2(x_0,0)$.
These results will be used to estimate breaking times for dispersive
fluids in section \ref{sec:dsw-breaking-time}.

\subsection{Viscous Shock Waves}
\label{sec:viscous-shock-waves}

It will be interesting to contrast the behavior of dispersive shock
waves for (\ref{eq:1}) with that of classical, viscous
shock waves resulting from a dissipative regularization of the
dispersionless equations.  For this, the jump and entropy conditions
for shocks are summarized below
\cite{courant_supersonic_1948,smoller_shock_1994}.

The Riemann problem (\ref{eq:77}) for (\ref{eq:121}) results in the
Hugoniot locus of classical shock solutions
\begin{equation}
  \label{eq:123}
  u_2 = u_1 \pm \left \{\frac{[P(\rho_2) - P(\rho_1)](\rho_2 -
      \rho_1)}{\rho_1 \rho_2} \right \}^{1/2} .
\end{equation}
The $-$ ($+$) corresponds to an admissible 1-shock (2-shock)
satisfying the Lax entropy conditions when the characteristic velocity
$\lambda_1$ ($\lambda_2$) \emph{decreases} across the shock so that
$\rho_2 > \rho_1 > 0$ ($\rho_1 > \rho_2 > 0$).  Weak 1-shocks
connecting the densities $\rho_1$ and $\rho_2 = \rho_1 + \Delta$,
$0 < \Delta \ll \rho_1$ exhibit the shock speed
\begin{equation}
  \label{eq:98}
  v^{(1)} \sim u_1 - c_1 - \frac{1}{2} \left ( \frac{c_1}{\rho_1} + c_1'
  \right ) \Delta, \qquad 0 < \Delta \ll \rho_1 .
\end{equation}
While the Riemann invariant $r_2$ exhibits a jump across the 1-shock,
it is third order in $\Delta/\rho_1$ so is approximately conserved for
weak shocks.  Weak, steady (non-propagating) 1-shocks satisfy the jump
conditions
\begin{eqnarray}
  \label{eq:127}
  \eqalign{
  \Delta \sim \frac{2 \rho_1 c_1}{c_1 + \rho_1 c_1'} (M_1 - 1), \\ 
  M_2 \sim 1 - (M_1 - 1), \qquad 0 < M_1 - 1 \ll 1,}
\end{eqnarray}
where $M_j = |u_j|/c_j$ are the Mach numbers of the up/downstream flows
and $c_1' \equiv \frac{d c}{d \rho}(\rho_1)$.  The upstream flow
indexed by $1$ is supersonic and the downstream flow is subsonic, this
behavior also holding for arbitrary amplitude shocks.

\subsection{Rarefaction Waves}
\label{sec:rarefaction-waves}

Centered rarefaction wave solutions of (\ref{eq:121}) exhibit the
following wave curves connecting the left and right states 
\numparts
\begin{eqnarray}
  \label{eq:254}
  \mathrm{1-rarefaction:}\quad u_1 &= u_2 + \int_{\rho_1}^{\rho_2}
  \frac{c(\rho)}{\rho} \rmd \rho, \qquad \rho_1 > \rho_2, \\
  \label{eq:255}
  \mathrm{2-rarefaction:}\quad u_1 &= u_2 - \int_{\rho_1}^{\rho_2}
  \frac{c(\rho)}{\rho} \rmd \rho , \qquad \rho_2 > \rho_1 ,
\end{eqnarray}
\endnumparts
where admissibility is opposite to the shock wave case.  The
characteristic velocities $\lambda_j$ \emph{increase} across a
rarefaction wave.  Since rarefaction waves are continuous and do not
involve breaking, the leading order behavior of dispersive and
dissipative regularizations for (\ref{eq:121}) are the same.  A
dispersive regularization of KdV
\cite{novikov_theory_1984,leach_large-time_2008} shows the development
of small amplitude oscillations for the first order singularities at
either the left or right edge of the rarefaction wave with one
decaying as $\mathcal{O}(t^{-1/2})$ and the other
$\mathcal{O}(t^{-2/3})$.  The width of these oscillations expands as
$\mathcal{O}(t^{1/3})$ \cite{gurevich_nonstationary_1974} so that
their extent vanishes relative to the rarefaction wave expansion with
$\mathcal{O}(t)$.

\subsection{Shock Tube Problem}
\label{sec:viscous-shock-tube}

Recall that the general solution of the Riemann problem consists of
three constant states connected by two waves, each either a
rarefaction or shock \cite{lax_hyperbolic_1973,smoller_shock_1994}.
The shock tube problem \cite{liepmann_elements_1957} involves a jump
in density for a quiescent fluid $u_1 = u_2 = 0$.  The solution
consists of a shock and rarefaction connected by a constant,
intermediate state $(\rhom,\um)$.  For the case $\rho_1 < \rho_2$, a
1-shock connects to a 2-rarefaction via the Hugoniot locus
(\ref{eq:123}) (with $-$) and the wave curve (\ref{eq:255}),
respectively.  For example, a polytropic gas with $P(\rho) = \kappa
\rho^{\gamma}$ gives the two equations
\begin{eqnarray}
  \label{eq:297}
  \eqalign{
  \mathrm{1-shock:} \qquad \um = -\left [\frac{(\kappa \rhom^\gamma -
      \kappa \rho_1^\gamma)(\rhom - 
      \rho_1)}{\rhom \rho_1} \right ]^{1/2}, \\
  \mathrm{2-rarefaction:} \qquad \um = - \frac{2 (\kappa
    \gamma)^{1/2}}{\gamma - 3} \left [ \rho_2^{(\gamma-1)/2} -
    \rhom^{(\gamma-1)/2} \right ] .}
\end{eqnarray}
Equating these two expressions provides an equation for the
intermediate density $\rhom$ and then the intermediate velocity $\um$
follows.

\section{Background: Simple DSWs}
\label{sec:normal-dsws}

The long time behavior of a DSW for the dispersive Euler model
(\ref{eq:1}) was first considered in \cite{el_resolution_2005}.  In
this section, the general Whitham-El construction of a simple wave led
DSW for step initial data is reviewed.  This introduces necessary
notation and background that will be used in the latter sections of
this work.

Analogous to the terminology for classical shocks, a 1-DSW is
associated with the $\lambda_1$ characteristic family of the
dispersionless system (\ref{eq:121}) involving left-going waves.  In
this case, the DSW leading edge is defined to be the leftmost (most
negative) edge whereas the DSW trailing edge is the rightmost edge,
these roles being reversed for the 2-DSW associated with the
$\lambda_2$ characteristic family.  There is a notion of polarity
associated with a DSW corresponding to its limiting behavior at the
leading and trailing edges.  The edge where the amplitude of the DSW
oscillations vanish (the harmonic limit) is called the linear wave
edge.  The soliton edge is associated with the phase boundary where
the DSW wavenumber $k \to 0$ (the soliton limit).  Thus, the soliton
edge could be the leading or trailing edge of the DSW, each case
corresponding to a different DSW polarity.  The polarity is generally
determined by admissibility criteria and typically follows directly
from the sign of the dispersion \cite{gurevich_nonlinear_1990}, as
will be shown in section \ref{sec:admiss-crit}.  The DSW construction
for 1-DSWs is outlined below.  A similar procedure holds for 2-DSWs.

Assuming the existence of a DSW oscillatory region described by slow
modulations of the periodic traveling wave from assumption A4, three
independent conservation laws are averaged with the periodic wave.
The wave's parameters $\rhobar$, the average density, $\ubar$, the
average velocity, $k$, the generalized (nonlinear) wavenumber, and
$a$, the wave amplitude are assumed to vary slowly in space and time.
The averaging procedure produces three first order, quasilinear PDEs.
This set combined with the conservation of waves, $k_t + \omega_x = 0$
($\omega$ here is the generalized, nonlinear frequency), following
from consistency of wave modulations, results in a closed system for
the modulation parameters, the Whitham modulation equations.  As
originally formulated by Gurevich and Pitaevskii
\cite{gurevich_nonstationary_1974}, the DSW free boundary value
problem is to solve the dispersionless equations (\ref{eq:121})
outside the oscillatory region and match this behavior to the
\emph{averaged} variables $\rhobar$ and $\ubar$ from the Whitham
equations at the interfaces with the oscillatory region where $k \to
0$ (soliton edge) or $a \to 0$ (linear wave edge).  These GP matching
conditions correspond to the coalescence of two characteristics of the
Whitham system at each edge of the DSW.  Assumption A5 can be used to
construct a self-similar, simple wave solution of the modulation
equations connecting the $k \to 0$ soliton edge with the $a \to 0$
linear wave edge via an integral curve so that the two DSW boundaries
asymptotically move with constant speed, the speeds of the double
characteristics at each edge.  In the Whitham-El method, the speeds
are determined by the following key mathematical observations
\cite{el_resolution_2005}
\begin{itemize}
\item The four Whitham equations admit \emph{exact} reductions to
  quasi-linear systems of three equations in the $k \to 0$ (soliton
  edge) and $a \to 0$ (linear wave edge) regimes.
\item Assuming a simple wave solution of the \emph{full} Whitham
  equations, one can integrate across the DSW with explicit knowledge
  only of the reduced systems in the $a = 0$ or $k = 0$ planes of
  parameters, thereby obtaining the DSW leading and trailing edge
  speeds.
\end{itemize}
This DSW closure method is appealing because it bypasses the difficult
determination and solution of the full Whitham equations.
Furthermore, it applies to a large class of nonintegrable nonlinear
wave equations.  Some nonintegrable equations studied with this method
include dispersive Euler equations like ion-acoustic plasmas
\cite{el_resolution_2005}, the Serre equations with zero surface
tension \cite{el_unsteady_2006}, the gNLS equation with
photorefractive \cite{el_theory_2007} and cubic/quintic nonlinearity
\cite{crosta_whitham_2012}, and other equations including the
Miyata-Camassa-Choi equations of two-layer fluids
\cite{esler_dispersive_2011}.

Simple DSWs are described by a simple wave solution of the Whitham
modulation system which necessitates self-similar variation in only
one characteristic field.  Using a nontrivial backward characteristic
argument, it has been shown that a simple wave solution requires the
constancy of one of the Riemann invariants (\ref{eq:17}) evaluated at
the left and right states \cite{el_resolution_2005}.  Then a necessary
condition for a simple DSW is one of 
\begin{eqnarray}
  \label{eq:20}
  \mathrm{1-DSW:} \quad u_2 = u_1 - \int_{\rho_1}^{\rho_2}
  \frac{c(\rho)}{\rho} \rmd \rho , \qquad \rho_2 > \rho_1 ,\\
  \label{eq:136}
  \mathrm{2-DSW:} \quad u_2 = u_1 + \int_{\rho_1}^{\rho_2}
  \frac{c(\rho)}{\rho} \rmd \rho , \qquad \rho_1 > \rho_2 .
\end{eqnarray}
\endnumparts
1-DSWs (2-DSWs) are associated with constant $r_2$ ($r_1$) hence vary
in the $\lambda_1$ ($\lambda_2$) characteristic field.  Equations
(\ref{eq:20}), (\ref{eq:136}) can be termed \emph{DSW loci} as they
are the dispersive shock analogues of the Hugoniot loci (\ref{eq:123})
for classical shock waves.  It is worth pointing out that the DSW loci
correspond precisely to the rarefaction wave curves in (\ref{eq:254})
and (\ref{eq:255}).  However, the admissibility criteria for DSWs
correspond to \emph{inadmissible}, compressive rarefaction waves where
the dispersionless characteristic speed decreases across the DSW.  The
coincidence of rarefaction and shock curves does occur in classical
hyperbolic systems but is restricted to a specific class, the
so-called \textit{Temple systems} \cite{temple_systems_1983} to which
the dispersionless Euler equations do not belong.

Recall from section \ref{sec:viscous-shock-waves} that across a viscous
shock, a Riemann invariant is conserved to third order in the jump
height.  Since the DSW loci (\ref{eq:20}), (\ref{eq:136}) result from
a constant Riemann invariant across the DSW, the DSW loci are equal to
the Hugoniot loci (\ref{eq:123}) up to third order in the jump
height.

\subsection{Linear Wave Edge}
\label{sec:linear-wave-edge-1}

The integral curve of the Whitham equations in the $a = 0$ (linear
wave edge) plane of parameters reduces to the relationships $k =
k(\rhobar)$, $\ubar = \ubar(\rhobar)$ and the ODE
\begin{equation}
  \label{eq:69}
  \frac{\rmd k}{\rmd \rhobar} = \frac{\omega_{\rhobar}}{\ubar(\rhobar) -
    c(\rhobar) - \omega_k} ,
\end{equation}
where the average velocity is constrained by the density through a
generalization of (\ref{eq:20})
\begin{equation}
  \label{eq:152}
  \ubar(\rhobar) = u_1 - \int_{\rho_1}^{\rhobar} \frac{c(\rho')}{\rho'} \rmd
  \rho' .
\end{equation}
The negative branch of the linear dispersion relation in
(\ref{eq:15}), $\omega = \ubar(\rhobar) k - \omega_0(k,\rhobar)$, is
associated with left-going waves, hence a 1-DSW.  
Using (\ref{eq:152}) and (\ref{eq:15}), (\ref{eq:69}) simplifies to
\begin{equation}
  \label{eq:197}
  \frac{\rmd k}{\rmd \rhobar} = \frac{c k /\rhobar +
     \omega_{0_{\rhobar}}}{c - \omega_{0_k} } .
\end{equation}
Equation (\ref{eq:197}) assumes $a = 0$, an exact reduction of the
Whitham equations only at the linear wave edge.  Global information
associated with the simple wave solution of the full Whitham equations
is obtained from the GP matching condition at the soliton edge by
noting that the modulation variables satisfy $(k,\rhobar,\ubar,a) =
(0,\rho_j,u_j,a_j)$ for some $j \in \{1,2\}$ depending on whether the
soliton edge is leading or trailing, \emph{independent} of the soliton
amplitude $a_j$. Thus (\ref{eq:197}) can be integrated in the $a = 0$
plane with the initial condition $k(\rho_j) = 0$, $\ubar(\rho_j) =
u_j$ to $k(\rho_{3-j})$ associated with the linear wave edge, giving
the wavenumber of the linear wave edge oscillations.  This
wavepacket's speed is then determined from the group velocity
$\omega_k$.

Based on the Riemann data (\ref{eq:77}), the integration domain for
(\ref{eq:197}) is $\rhobar \in [\rho_1,\rho_2]$.  The initial
condition occurs at either the leading edge where $\rhobar = \rho_1$
or the trailing edge where $\rhobar = \rho_2$.  The determination of
the location of the linear wave edge, leading or trailing, is set by
appropriate admissibility conditions discussed in
section \ref{sec:admiss-crit}.  The solution of (\ref{eq:197}) with
initial condition at $\rho_j$ will be denoted $k(\rhobar;\rho_j)$ so
that one of
\begin{equation}
  \label{eq:73}
  k(\rho_j;\rho_j) = 0, \qquad
  j = 1, 2 ,
\end{equation}
holds for (\ref{eq:197}).  Evaluating the solution of
(\ref{eq:197}) at the linear wave edge $k(\rho_j;\rho_{3-j})$, gives
the wavenumber of the linear wavepacket at the leading (trailing) edge
when $j = 1$ ($j = 2$).  The associated 1-DSW linear wave edge speed is
denoted $v_j(\rho_1,\rho_2)$ and is found from the group velocity
evaluated at $k(\rho_j;\rho_{3-j})$
\begin{eqnarray}
  \label{eq:75}
  \eqalign{
  v_j(\rho_1,\rho_2) &=
  \omega_k[k(\rho_j;\rho_{3-j}),\rho_{j}] \\
  &= \ubar(\rho_{j}) -
  \omega_{0_k}[k(\rho_{j};\rho_{3-j}),\rho_{j}] , \qquad j = 1,2.}
\end{eqnarray}

\subsection{Soliton Edge}
\label{sec:soliton-edge}

An exact description of the soliton edge where $k\to 0$ involves the
dispersionless equations for the average density $\rhobar$ and
velocity $\ubar$ as well as an equation for the soliton amplitude $a$.
While, in principle, one can carry out the simple DSW analysis on
these equations, it is very convenient to introduce a new variable
$\kt$ called the conjugate wavenumber.  It plays the role of an
amplitude and depends on $\rhobar$, $\ubar$, and $a$ thus is not a new
variable.  The resulting modulation equations at the soliton edge then
take a correspondingly symmetric form relative to the linear edge.
The speed of the soliton edge is determined in an analogous way to the
linear edge by introducing a conjugate frequency
\begin{equation}
  \label{eq:26}
  \eqalign{
    \tilde{\omega}(\tilde{k},\rhobar) &= -\rmi
    \omega(\rmi\tilde{k},\rhobar) = \ubar(\rhobar) \tilde{k} + \rmi
    \omega_0(\rmi \tilde{k},\rhobar) \\
    &= \ubar(\rhobar) \tilde{k} + 
    \omegat_0(\tilde{k},\rhobar) . 
  }
\end{equation}
The conjugate wavenumber plays the role analogous to an amplitude so that $\kt
\to 0$ corresponds to the linear wave edge where $a \to 0$.
Integrating the ODEs for a simple wave in the $k = 0$ plane results in
\begin{equation}
  \label{eq:3}
  \frac{\rmd \kt}{\rmd \rhobar} = \frac{c(\rhobar) \kt /\rhobar +
     \omegat_{0_{\rhobar}}}{c(\rhobar) - \omegat_{0_{\kt}} } ,
\end{equation}
the same equation as (\ref{eq:69}) but with conjugate variables.  It
is remarkable that the description of the soliton edge so closely
parallels that of the linear wave edge.  The initial condition is
given at the linear wave edge where $\kt = 0$.  As in (\ref{eq:73}),
the solution with zero initial condition at $\rhobar = \rho_j$ is
denoted $\kt(\rhobar;\rho_j)$ according to
\begin{equation}
  \label{eq:53}
  \kt(\rho_j;\rho_j) = 0 , \qquad j = 1,2.
\end{equation}
Then the soliton speed $s_j$, $j = 1, 2$ is the conjugate phase
velocity evaluated at the conjugate wavenumber associated with the
soliton edge
\begin{eqnarray}
  \label{eq:25}
  \eqalign{
  s_j(\rho_1,\rho_2) &=
  \frac{\omegat[\kt(\rho_j;\rho_{3-j}),\rho_j]}{\tilde{k}(\rho_j;\rho_{3-j})}
  \\ 
  &= \ubar(\rho_j) -
  \frac{\omegat_0[\kt(\rho_j;\rho_{3-j}),\rho_j]}{\kt(\rho_j;\rho_{3-j})} .}
\end{eqnarray}

\begin{remark}
  \label{sec:soliton-edge-1}
  In a number of example dispersive fluids studied in this work and
  elsewhere, the transformation to the scaled phase speed 
  \begin{equation*}
    \alpha(\rhobar) = \frac{\omega_0[k(\rhobar),\rhobar]}{c(\rhobar)
      k(\rhobar)} ,
  \end{equation*}
  of the dependent variable in
  (\ref{eq:197}) and the analogous transformation
  \begin{equation*}
    \tilde{\alpha}(\rhobar) = \frac{\omegat_0[\kt(\rhobar),\rhobar]}{c(\rhobar)
      \kt(\rhobar)} ,
  \end{equation*}
  for (\ref{eq:3}) are helpful, reducing the ODEs (\ref{eq:197}) and
  (\ref{eq:3}) to simpler and, often, separable equations for $\alpha$
  and $\alphat$.
\end{remark}

\subsection{Dispersive Riemann Problem}
\label{sec:disp-riem-probl}

The integral wave curves (\ref{eq:254}), (\ref{eq:255}) and the DSW
loci (\ref{eq:20}), (\ref{eq:136}) can be used to solve the dispersive
Riemann problem (\ref{eq:77}) just as the wave curves and the Hugoniot
loci are used to solve the classical Riemann problem
\cite{smoller_shock_1994}.  In both cases, the solution consists of
two waves, one for each characteristic family, connected by a constant
intermediate state.  Each wave is either a rarefaction or a shock.

\begin{figure}
  \centering
  \includegraphics[scale=1]{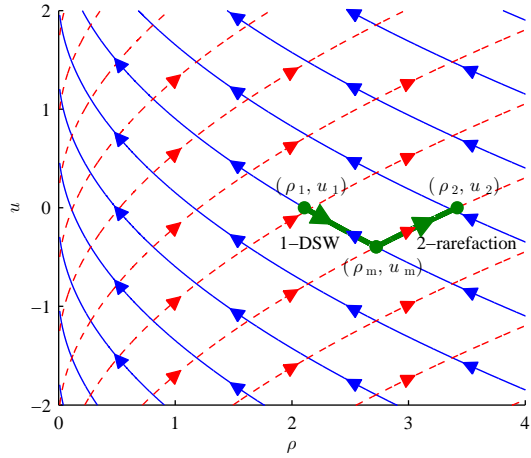}
  \caption{Integral curves and DSW loci for the dispersive Euler
    equations with $c(\rho) = \sqrt{\rho}$.}
  \label{fig:integral_curves}
\end{figure}

In contrast to the classical case, the integral wave curves and the
DSW loci are the \emph{same} for the dispersive Riemann problem.  It
is the direction in which they are traversed that determines
admissibility of a rarefaction or a DSW.  This enables a convenient,
graphical description of solutions to the dispersive Riemann problem
as shown in figure \ref{fig:integral_curves}.  Solid curves
({\color{blue}\full}) correspond to example 1-wave curves 
(\ref{eq:20}), (\ref{eq:254}) and the dashed curves ({\color{red}\dashed})
correspond to example 2-wave curves (\ref{eq:136}), (\ref{eq:255}).
The arrows provide the direction of increasing dispersionless
characteristic speed for each wave family.  Tracing an integral curve
in the direction of increasing characteristic speed corresponds to an
admissible rarefaction wave.  The decreasing characteristic speed
direction corresponds to an admissible DSW.  The solution to a
dispersive Riemann problem is depicted graphically by tracing
appropriate integral curves to connect the left state $(\rho_1,u_1)$
with the right state $(\rho_2,u_2)$.  There are multiple paths
connecting these two states but only one involves two admissible
waves.  This is shown by the thick curve in figure
\ref{fig:integral_curves}.  Since the 1-wave curve is traced in the
negative direction to the intermediate, constant state $(\rhom,\um)$,
this describes a 1-DSW.  The 2-wave curve is then traced in the
positive direction to the right state, describing a 2-rarefaction.
The 1-DSW is admissible because $\rhom > \rho_1$.  Since the
characteristic speed $\lambda_2$ is monotonically increasing, the
2-rarefaction is admissible ($\rhom < \rho_2$).  The direction of the
curve connecting the left and right states was taken from left to
right.  The opposite direction describes an inadmissible 1-rarefaction
connected to an inadmissible 2-DSW.

The example shown in figure \ref{fig:integral_curves} corresponds to the
dispersive shock tube problem consisting of an arbitrary jump in
density for a quiescent fluid $u_1 = u_2 = 0$.  Such problems have
been studied in a number of dispersive fluids,
e.g.~\cite{el_decay_1995,el_unsteady_2006,khamis_nonlinear_2008,esler_dispersive_2011}.
The determination of $(\rhom,\um)$ proceeds by requiring that the left
state $(\rho_1,0)$ lie on the 1-DSW locus (\ref{eq:20})
\begin{equation}
  \label{eq:277}
  \um = - \int_{\rho_1}^{\rhom} \frac{c(\rho)}{\rho} \rmd \rho .
\end{equation}
The second wave connects $(\rhom,\um)$ to the right state
$(\rho_2,0)$ via the 2-rarefaction wave curve (\ref{eq:255})
\begin{equation}
  \label{eq:293}
  \um = -\int_{\rhom}^{\rho_2} \frac{c(\rho)}{\rho} \rmd \rho .
\end{equation}
Equating (\ref{eq:277}) and (\ref{eq:293}) leads to
\begin{equation*}
  \int_{\rho_1}^{\rhom} \frac{c(\rho)}{\rho} \rmd \rho -
  \int_{\rhom}^{\rho_2} \frac{c(\rho)}{\rho} \rmd \rho = 0 ,
\end{equation*}
which determines the intermediate density $\rho_1 < \rhom < \rho_2$.
The intermediate velocity $\um < 0$ follows.  

This construction of the wave types and the intermediate state
$(\rhom,\um)$ is independent of the sign of dispersion and the details
of the dispersive term $D$ in (\ref{eq:1}), depending only upon the
pressure law $P(\rho)$.  For example, polytropic dispersive fluids
with $P(\rho) = \kappa \rho^\gamma$ (e.g., gNLS with power law
nonlinearity (\ref{eq:164}) and gSerre (\ref{eq:96})) yield the
intermediate state (previously presented in \cite{gurevich_expanding_1984})
\begin{eqnarray}
  \label{eq:295}
  \eqalign{
  \rhom = \left [ \frac{1}{2} \left ( \rho_1^{(\gamma-1)/2} +
      \rho_2^{(\gamma - 1)/2} \right ) \right
  ]^{2/(\gamma-1)} , \\
  \um = \frac{2 (\kappa
    \gamma)^{1/2}}{3 - \gamma} \left [ \rho_1^{(\gamma-1)/2} -
    \rho_2^{(\gamma-1)/2} \right ] .}
\end{eqnarray}
This prediction will be compared with numerical computations of gNLS
in section \ref{sec:polytropic-gas-frho}.

\section{DSW Admissibility Criteria}
\label{sec:admiss-crit}

As shown in section \ref{sec:rarefaction-waves}, when (\ref{eq:20})
holds and $\rho_1 > \rho_2$, a continuous 1-rarefaction wave solution
to the dispersionless equations exists.  Gradient catastrophe does not
occur so the rarefaction wave correctly captures the leading order
behavior of the dispersive regularization.  However, when $\rho_1 <
\rho_2$, the rarefaction wave solution is no longer admissible and
integrating the dispersionless equations via the method of
characteristics results in a multivalued solution.  A dispersive
regularization leading to a DSW is necessitated.  Specific criteria
are now provided to identify admissible 1-DSWs.

The general admissibility criteria for DSWs depend on the ordering of
the soliton and linear wave edges.  In what is termed here a
``$1^+$-DSW'', the conditions are \cite{el_resolution_2005}
\numparts
\begin{eqnarray}
  \label{eq:283}
  &u_2 - c_2 < s_2 <
  u_2 + c_2, \\ 
  \label{eq:284}
  1^+\mathrm{-DSW:} \qquad &v_1 < 
  u_1 - c_1, \\
  \label{eq:68}
  &s_2 > v_1 ,
\end{eqnarray}
\endnumparts
where the soliton is at the \emph{trailing} edge of the DSW.
Similarly a ``$1^-$-DSW'' satisfies the conditions \numparts
\begin{eqnarray}
  \label{eq:281}
  &u_2 - c_2 < v_2 <
  u_2 + c_2, \\
  \label{eq:70}
  1^-\mathrm{-DSW:} \qquad & 
  s_1 < 
  u_1 - c_1, \\
  \label{eq:282}
  &v_2 > s_1 ,
\end{eqnarray}
\endnumparts
with the soliton at the \emph{leading} edge.  The designation
$1^+$-DSW ($1^-$-DSW) corresponds to a positive (negative) dispersion
fluid as shown below.  Recall that the soliton (linear) edge speed
$s_j$ ($v_j$) corresponds to the left edge of the DSW if $j=1$ or the
right edge if $j=2$.  Thus, the subscript determines the polarity of
the DSW.  These conditions are analogous to the Lax entropy conditions
for dissipatively regularized hyperbolic systems
\cite{lax_hyperbolic_1973}.  A key difference with classical fluids is
that there is only one ``sign'' of dissipation due to time
irreversibility.  For time-reversible, dispersive fluids, both signs
are possible.  The admissibility criteria ensure that only three
characteristics impinge upon the three parameter DSW, transferring
initial/boundary data into the DSW and allowing for the simple wave
condition (\ref{eq:20}) to hold.  Sufficient conditions for these
criteria as applied to the dispersive Euler equations are now shown.

First, consider the criteria for a $1^+$-DSW in a positive dispersion
fluid.  Inserting the linear wave and soliton edge speeds
(\ref{eq:75}), (\ref{eq:25}) into inequalities (\ref{eq:283}),
(\ref{eq:284}) simplifies the first two admissibility criteria to
\numparts
\begin{eqnarray}
  \label{eq:270}
  -c_2 < \frac{\omegat_0(\kt_2,\rho_2)}{\kt_2} < c_2 , \\
  \label{eq:287}
  \omega_{0_k}(k_1,\rho_1) > c_1 ,
\end{eqnarray}
\endnumparts
where $\kt_2 = \kt(\rho_2;\rho_1)$ and $k_1 = k(\rho_1;\rho_2)$.  In
section \ref{sec:admissibility}, the admissibility of weak $1^+$-DSWs
when $0 < \rho_2 - \rho_1 \ll 1$ is demonstrated.  The further
assumptions
\begin{equation}
  \fl
  \label{eq:307}
  1^+\mathrm{-DSW:} \qquad \omegat_{{\kt\kt}} < 0, \quad
  c k + \rhobar \omega_{0_{\rhobar}} > 0,  \quad
  c \kt + \rhobar \omegat_{0_{\rhobar}} > 0,  
\end{equation}
enable the extrapolation of Eulerian $1^+$-DSW admissibility to
moderate and large jumps $\rho_2 > \rho_1$ (below).  Assumptions
(\ref{eq:307}) hold for gNLS, gSerre, and ion-acoustic plasma in
certain parameter regimes (moderate jumps).

The extrapolation of admissibility to larger jumps can be demonstrated
as follows.  For 1-DSWs, the negative branch of the dispersion
relation (\ref{eq:15}) has been chosen for $k > 0$.  Using the small
$k$ asymptotics (\ref{eq:80}) in (\ref{eq:197}) with initial condition
$k(\rho_2;\rho_2) = 0$, one can show that $k(\rhobar;\rho_2)$ is a
decreasing function of $\rhobar$ for $ | \rhobar - \rho_2 | \ll
\rho_2$.  Since $\omega_{0_k}(0,\rhobar) = c(\rhobar)$, the convexity
of $\omega_0$ implies $\omega_{0_k}(k,\rhobar) > c(\rhobar)$ for $k >
0$.  This fact combined with (\ref{eq:307}) in (\ref{eq:197}) implies
that $k_1 = k(\rho_1;\rho_2) > 0$ for $\rho_2 > \rho_1$ and that
inequality (\ref{eq:287}) holds.  Similar arguments demonstrate that
$\kt_2 = \kt(\rho_2;\rho_1) > 0$ for $\rho_2 > \rho_1$ and that the
inequalities in (\ref{eq:270}) hold.  Thus, if (\ref{eq:307}) hold for
an interval $k \in [0,k_*)$, then (\ref{eq:270}) and (\ref{eq:287})
are verified for $\rho_2 \in (\rho_1,\rho_*)$ where $k(\rho_1;\rho_*)
= k_*$.  It is now clear why $\rho_2 > \rho_1$ in (\ref{eq:20}) and
the designation $1^+$-DSW is used when the sign of dispersion is
positive.

By similar arguments, inequalities (\ref{eq:281}) and
(\ref{eq:70}) hold for negative dispersion fluids when $\rho_1 <
\rho_2$ and
\begin{eqnarray}
  \label{eq:82}
  \fl
  1^-\mathrm{-DSW:} \qquad \omegat_{{\kt\kt}} > 0, \quad
  c k + \rhobar \omega_{0_{\rhobar}} > 0,  \quad
  c \kt + \rhobar \omegat_{0_{\rhobar}} > 0 .
\end{eqnarray}
The only change with respect to (\ref{eq:307}) is the convexity of the
conjugate dispersion relation.
 
The final inequalities (\ref{eq:68}) and (\ref{eq:282}) require
verification.  An explicit analysis for weak DSWs is given in section
\ref{sec:weak-dsws}.  An intuitive argument can be given for the
general case.  When $\omega_{kk} > 0$, the case of positive
dispersion, the group velocity of waves with shorter wavelengths is
larger while the opposite is true for negative dispersion.  The
soliton edge corresponds to the longest wavelength (infinite) hence
should be the trailing (leading) edge in positive (negative)
dispersion systems.  For the 1-DSW, a trailing soliton edge means $s >
v$ as is the case in (\ref{eq:68}).  A leading soliton edge
corresponds to the ordering in (\ref{eq:282}).

In summary, sufficient conditions for a simple wave led $1$-DSW are
$\rho_1 < \rho_2$ and either (\ref{eq:68}), (\ref{eq:307}) for a
positive dispersion $1^+$-DSW or (\ref{eq:282}), (\ref{eq:82}) for a
negative dispersion $1^-$-DSW.  Because of this, it is convenient to
dispense with the subscripts defining the DSW speeds $v_j$ and $s_j$
in (\ref{eq:75}), (\ref{eq:25}) and use the notation
\begin{eqnarray}
  \label{eq:103}
  v_- &\equiv v_2, \qquad v_+ \equiv
  v_1, \qquad s_- &\equiv s_1, \qquad s_+ \equiv s_2 ,
\end{eqnarray}
which identifies the dispersion sign.  The conditions (\ref{eq:68}) or
(\ref{eq:282}) can be verified a-priori while the speed orderings
(\ref{eq:307}) or (\ref{eq:82}) must be verified by computing the
speeds directly.  For the case of $2^\pm$-DSWs, the requirement is
$\rho_1 > \rho_2$ and the appropriate ordering of the soliton and
linear wave edges.  The Lax entropy conditions for the dissipative
regularization of the Euler equations give similar criteria, namely
positive (negative) jumps for 1-shocks (2-shocks)
\cite{smoller_shock_1994}.

\subsection{Nonstationary and Stationary Soliton Edge}
\label{sec:stat-solit-edge}

Due to Galilean invariance, there is flexibility in the choice of
reference frame for the study of DSWs.  In the general construction of
$1$-DSWs presented here, the laboratory frame was used with the
requirement that the background flow variables lie on the 1-DSW locus
(\ref{eq:20}).  With this coordinate system, three of the four
background flow properties $(\rho_1,\rho_2,u_1,u_2)$ can be fixed
while the fourth is determined via the 1-DSW locus.  The soliton and
linear wave edge speeds follow according to (\ref{eq:25}) and
(\ref{eq:75}) such that either one of the admissibility criteria for a
$1^+$- or $1^-$-DSW hold.

Another convenient coordinate system is one moving with the soliton
edge as shown in figure \ref{fig:pos_neg_dispersion}.  In this case, one
can consider the upstream quantities $\rho_1 > 0$, $u_1 > 0$ given and
the stationary condition
\begin{equation}
  \label{eq:21}
  s_\pm(\rho_1,\rho_2) = 0,
\end{equation}
to hold.  The downstream density $\rho_2$ is computed from
(\ref{eq:21}) while the downstream velocity $u_2$ follows from the
1-DSW locus~(\ref{eq:20}).  The linear wave edge speed is found, as
usual, from (\ref{eq:75}).

The admissibility criteria (\ref{eq:283})--(\ref{eq:282}) for a 1-DSW
with stationary soliton edge become
\numparts
\begin{eqnarray*}
  \fl
  &1^+\mathrm{-DSW:}\qquad M_2 - 1 < 0 <
  M_2 + 1, \qquad
  \frac{v_1}{c_1}  < M_1 - 1, \qquad
  v_1 < 0 , \\
  \fl
  &1^-\mathrm{-DSW:}\qquad
  M_2 - 1 < \frac{v_2}{c_2} <
  M_2 + 1, \qquad
  1  < M_1, \qquad
  v_2 > 0 .
\end{eqnarray*}
\endnumparts
For $1^+$-DSWs, the downstream flow must be subsonic ($M_2 < 1$) while
for $1^-$-DSWs, the upstream flow must be supersonic ($M_1 > 1$).  It
is expected that both properties hold for both $1^\pm$-DSWs but this
is not required by the admissibility criteria.  Supersonic upstream
flow and subsonic downstream flow is consistent with classical shock
waves and will be demonstrated for weak DSWs in section
\ref{sec:weak-dsws}.

The DSW locus (\ref{eq:20}), linear wave edge speed in (\ref{eq:75}),
and the soliton edge speed in~(\ref{eq:25}) along with the stationary
condition (\ref{eq:21}) constitute the jump conditions for a 1-DSW
with a stationary soliton edge in dispersive Eulerian fluids.  Given
the upstream Mach number $M_1$ and density $\rho_1$, the stationary
condition (\ref{eq:21}) and the soliton edge speeds (\ref{eq:25})
determine the downstream density $\rho_2$ while the DSW locus
(\ref{eq:20}) determines the downstream Mach number $M_2$.

Figure \ref{fig:pos_neg_dispersion} depicts generic descriptions of
$1^\pm$-DSWs with stationary soliton edge.  In the $1^-$-DSW case of
figure \ref{fig:pos_neg_dispersion}(a), the soliton edge is the
leading edge and the linear wave edge is the trailing edge.  The
opposite orientation is true for the $1^+$-DSW shown in figure
\ref{fig:pos_neg_dispersion}(b).  This generic behavior was also
depicted in \cite{gurevich_nonlinear_1990} based on an analysis of
weak DSWs in plasma.

\subsection{Breakdown of Simple Wave Assumption}
\label{sec:breakd-simple-wave}

A fundamental assumption in this DSW construction is the existence of
a simple wave or integral curve of the full Whitham modulation
equations connecting the trailing and leading edge states.  This
assumption is in addition to the admissibility criteria discussed in
section \ref{sec:admiss-crit}.  The simple wave assumption for a 1-DSW
requires a monotonic decrease of the associated characteristic speed
as the integral curve is traversed from left to right
\cite{dafermos_hyperbolic_2009}.  This monotonicity condition leads to
the requirement of genuine nonlinearity of the full modulation system.
Identification of the loss of monotonicity is undertaken by examining
the behavior of the modulation system at the leading and trailing
edges.

The full Whitham modulation system exhibits four characteristic speeds
$\lambda_1 \le \lambda_2 \le \lambda_3 \le \lambda_4$ that depend on
$(k,\rhobar,\ubar,a)$.  In the case of positive dispersion, the simple
wave DSW integral curve is associated with the 2-characteristic
\cite{el_resolution_2005} and connects the left, right states
$(k_1,\rho_1,u_1,0)$, $(0,\rho_2,u_2,a_2)$, respectively where $k_1$
is the wavenumber of the wavepacket at the linear wave edge and $a_2$
is the solitary wave amplitude at the soliton edge.  Generally, the
integral curve can be parametrized by $\rhobar \in [\rho_1,\rho_2]$,
$(k(\rhobar),\rhobar,\ubar(\rhobar),a(\rhobar))$.  The monotonicity
condition for a simple wave can therefore be expressed as
\begin{equation}
  \label{eq:72}
  0 > \frac{\rmd \lambda_2}{\rmd \rhobar} = \pd{\lambda_2}{k} k' +
  \pd{\lambda_2}{\rhobar} + \pd{\lambda_2}{\ubar} \ubar' +
  \pd{\lambda_2}{a} a' ,
\end{equation}
where primes denote differentiation with respect to $\rhobar$ along
the integral curve.  At the linear wave edge, the $\lambda_2$ and
$\lambda_1$ characteristics merge
\begin{equation}
  \label{eq:35}
  \nu_1(k,\rhobar,\ubar) = \lim_{a \to 0^+}
  \lambda_1(k,\rhobar,\ubar,a) = \lim_{a\to 0^+}
  \lambda_2(k,\rhobar,\ubar,a) ,
\end{equation}
where $\nu_1$ is the smallest characteristic speed of the modulation
system when $a = 0$.  This merger of characteristics implies that
right differentiability of $\lambda_2$ when $a = 0$ requires
\begin{equation}
  \label{eq:74}
  \pd{}{a} \lambda_2(k,\rhobar,\ubar,0) = \pd{}{a}
  \lambda_1(k,\rhobar,\ubar,0) = 0 .
\end{equation}
Using (\ref{eq:72}), (\ref{eq:35}), and (\ref{eq:74}), the breakdown
of the monotonicity condition (\ref{eq:72}) can now be identified at
the linear wave edge as
\begin{equation}
  \label{eq:76}
  \fl
  \left . \lim_{a \to 0^+} \frac{\rmd \lambda_2}{\rmd \rhobar} \right
  |_{k=k_1,\rhobar=\rho_1,\ubar=u_1} = \left . \pd{\nu_1}{k} 
    k' + \pd{\nu_1}{\rhobar} + \pd{\nu_1}{\ubar} \ubar' \right
  |_{k=k_1,\rhobar=\rho_1,\ubar = u_1}   = 0 .
\end{equation}
The zero-amplitude reduction of the modulation system is comprised of
the two dispersionless equations and the conservation of waves
\cite{el_resolution_2005}
\begin{equation}
  \label{eq:299}
  \left [
    \begin{array}{c}
      \rhobar \\
      \ubar \\
      k
    \end{array}
  \right ]_t + 
  \left[
    \begin{array}{ccc}
      \ubar & \rhobar & 0 \\
      c^2/\rhobar & \ubar & 0 \\
      \omega_{{\rhobar}} & \omega_{\ubar} & \omega_{k}
    \end{array} 
  \right ]
  \left [
    \begin{array}{c}
      \rhobar \\
      \ubar \\
      k
    \end{array}
  \right ]_x ,
\end{equation}
where $\omega(k,\rhobar,\ubar)$ is the negative branch of the
dispersion relation (\ref{eq:15}).  The eigenvalues $\nu_{1,2,3}$ and
associated right eigenvectors $\bi{r}_{1,2,3}$ for this hyperbolic
system are \numparts
\begin{eqnarray}
  \label{eq:300}
  (\nu_1,\bi{r}_1) &= (\ubar - \omega_{0_k},[
  0,0,1]^T ) , \\
  \label{eq:141}
  (\nu_2,\bi{r}_2) &= (\ubar - c, [
  \rhobar(\omega_{0_k} - c),-c(\omega_{0_k}-c),ck - \rhobar
  \omega_{0_{\rhobar}} ]^T ), \\
  \label{eq:142}
  (\nu_3,\bi{r}_3) &= (\ubar + c, [
  \rhobar(\omega_{0_k} + c),c(\omega_{0_k}+c),ck - \rhobar
  \omega_{0_{\rhobar}} ]^T ) .
\end{eqnarray}
\endnumparts
Note the ordering $\nu_1 < \nu_2 < \nu_3$ due to the $1^+$-DSW
admissibility criterion (\ref{eq:287}).  Recalling that the 1-DSW
integral curve satisfies (\ref{eq:152}) and (\ref{eq:197}) at the
linear wave edge, then (\ref{eq:76}) occurs (breakdown) when
\begin{equation}
  \label{eq:137}
  \left . \omega_{0_{kk}} \left ( \omega_{0_{\rhobar}} + \frac{c k}{\rhobar} 
  \right ) + \left ( c - \omega_{0_k} \right ) \left (
    \omega_{0_{k\rhobar}} + \frac{c}{\rhobar} \right 
  ) \right |_{k = k_1, \rhobar = \rho_1}  = 0 .
\end{equation}
A direct computation shows that $\partial v_+/\partial \rho_1 = 0$ if
and only if (\ref{eq:137}) holds, offering a simple test for linear
degeneracy once the DSW speed has been computed.  The coalescence of
two characteristic speeds (non-strict hyperbolicity) implies linear
degeneracy \cite{dafermos_hyperbolic_2009}. DSWs described by
modulation systems lacking strict hyperbolicity and genuine
nonlinearity have been studied for integrable systems
\cite{marchant_undular_2006,pierce_self_2006,pierce_large_2007,pierce_self_2007,kodama_on_2008,grava_initial_2009,kamchatnov_undular_2012}.
The results indicate a number of novel features including compound
waves (e.g., a DSW attached to a rarefaction), kinks, and enhanced
curvature of the DSW oscillation envelope.  The latter has lead
previous authors \cite{kodama_on_2008} to describe such non-simple
DSWs as having a ``wineglass shape'' in contrast with simple DSWs that
exhibit a ``martini glass shape'' (see figure
\ref{fig:pos_neg_dispersion}).  The linear degeneracy condition
(\ref{eq:137}) was given in \cite{el_unsteady_2006} for the
non-integrable Serre equations.  Its derivation, however, was wrongly
attributed to loss of genuine nonlinearity in the reduced modulation
system (\ref{eq:299}) when $a = 0$.  Linear degeneracy occurs in this
system when
\begin{equation}
  \label{eq:144}
  \nabla \nu_i \cdot \bi{r}_i = 0 ,
\end{equation}
for some $i \in \{1,2,3\}$.  $1^+$-DSW admissibility (\ref{eq:287})
implies that the 2- and 3-characteristic fields do not exhibit linear
degeneracy.  Evaluation of (\ref{eq:144}) for the 1-characteristic
field, however, gives
\begin{equation}
  \label{eq:200}
  \omega_{{kk}}(k_1,\rho_1) = 0,
\end{equation}
corresponding to zero dispersion, a different condition than
(\ref{eq:137}).  In the vicinity of the trailing edge, the $1^+$-DSW
self-similar simple wave corresponds to the first characteristic
family (\ref{eq:300}) and satisfies the ODEs
\begin{equation*}
  \rhobar' = 0, \qquad \ubar' = 0, \qquad k' = -1/\omega_{0_{kk}} ,
\end{equation*}
where differentiation is with respect to the self-similar variable
$x/t$.  This demonstrates that the Whitham modulation equations
exhibit gradient catastrophe, $|k'| \to \infty$, when the dispersion
is zero (\ref{eq:200}).  A direct computation demonstrates that
$\partial v_+/\partial \rho_2 = 0$ if and only if (\ref{eq:200})
holds.  Thus, breaking in the Whitham modulation equations coincides
with an extremum of the linear wave edge speed with respect to
variation in $\rho_2$.

Breaking in the Whitham equations has been resolved in specific
systems by appealing to modulated multiphase waves describing DSW
interactions
\cite{el_generating_1996,grava_generation_2002,biondini_whitham_2006,hoefer_interactions_2007,ablowitz_soliton_2009}.
A recent study of DSWs in the scalar magma equation shows that the
development of zero dispersion for single step initial data leads to
internal multiphase dynamics termed DSW implosion
\cite{lowman_dispersive_2013}.  The simple wave assumption no longer
holds.  This behavior was intuited by Whitham before the development
of DSW theory (see \cite{whitham_linear_1974}, section 15.4) where
breaking of the Whitham modulation equations were hypothesized to
``represent a source of oscillations''.

An analysis of the soliton wave edge where $k \to 0$ can be similarly
undertaken.  Recalling that the characteristic speed of the soliton
edge is the phase velocity $\ubar - \omegat_0/\kt$ (\ref{eq:25}), the
breakdown of the monotonicity condition for the positive dispersion
case is
\begin{equation*}
  \fl
  \left . \lim_{k \to 0^+} \frac{\rmd \lambda_2}{\rmd \rhobar} \right
  |_{\kt = \kt_2,\rhobar = \rho_2,\ubar=u_2} =
  \left . -\frac{\partial}{\partial \kt} \left ( \frac{\omegat_0}{\kt} \right )
  \kt' - \frac{\partial}{\partial \rhobar} \left (
    \frac{\omegat_0}{\kt} \right ) + \ubar' \right |_{\kt =
  \kt_2,\rhobar = \rho_2,\ubar=u_2} = 0 .
\end{equation*}
Using the $1$-DSW locus (\ref{eq:20}) and the characteristic ODE
(\ref{eq:3}) lead to the simplification
\begin{equation*}
  \left . \left ( \kt c - \omegat_0 \right ) \left ( \omegat_{0_{\rhobar}} +
    \frac{c \kt}{\rhobar} \right ) \right |_{\kt = \kt_2,\rhobar =
  \rho_2} = 0 .
\end{equation*}
The positivity of the first factor is equivalent to the admissibility
criterion (\ref{eq:270}) so it is a zero of the second factor
\begin{equation}
  \label{eq:100}
  \left . \omegat_{0_{\rhobar}} + \frac{c \kt}{\rhobar} \right |_{\kt
    = \kt_2,\rhobar = \rho_2} = 0 ,
\end{equation}
that offers a new route to linear degeneracy.  Recalling that the
dispersion relation involves two branches (\ref{eq:15}), care must be
taken that the appropriate branch is used in (\ref{eq:100}), which can
change when passing through $\omegat_0 = 0$.  A direct computation
verifies that $\partial s_+/\partial \rho_2 = 0$ if and only if
(\ref{eq:100}) holds.  Therefore, an easy test for linear degeneracy
is to find an extremum of $s_+(\rho_1,\rho_2)$ with respect to
variations in $\rho_2$.  Note that the linear degeneracy condition
(\ref{eq:100}) also coincides with a breaking of one of the additional
sufficient admissibility conditions in (\ref{eq:307}).

Just as zero dispersion at the linear wave edge can lead to
singularity formation in the Whitham equations, the soliton edge can
similarly exhibit catastrophe when the phase velocity reaches an
extremum
\begin{equation}
  \label{eq:302}
  \left . \left ( \frac{\omegat_0}{\kt} \right )_{\kt}
  \right |_{\kt = \kt_2,\rhobar = \rho_2} = 0 .
\end{equation}
This corresponds to zero conjugate dispersion.  When (\ref{eq:302}) is
satisfied, wave interactions at the leading edge are expected to occur
for larger initial jumps.  In contrast to the linear degeneracy
criterion, a direct computation verifies that $\partial s_+/\partial
\rho_1 = 0$ if and only if (\ref{eq:302}) holds.

The criteria for breakdown of $1^-$-DSWs is the same as
(\ref{eq:137}), (\ref{eq:200}) with $1 \to 2$ and (\ref{eq:100}),
(\ref{eq:302}) with $2 \to 1$.  In summary, two mechanisms at each DSW
edge for the breakdown of the simple wave assumption have been
identified: the loss of monotonicity (linear degeneracy)
(\ref{eq:137}), (\ref{eq:100}) or gradient catastrophe in the Whitham
modulation equations due to zero dispersion (\ref{eq:200}),
(\ref{eq:302}).  These behaviors can be succinctly identified via
extrema in the DSW speeds as
\numparts
\begin{eqnarray}
  \label{eq:106}
  \fl
  &1^+\mathrm{-DSW:}\qquad
  \begin{array}{ll}
    \mbox{linear degeneracy} & {\displaystyle \frac{\partial
        v_+}{\partial \rho_1} = 
      0  \quad \mathrm{or} \quad  \frac{\partial
        s_+}{\partial \rho_2} = 
    0} , \\
    \mbox{gradient catastrophe} & {\displaystyle \frac{\partial
        v_+}{\partial \rho_2} = 
    0 \quad \mathrm{or} \quad \frac{\partial s_+}{\partial \rho_1} =
    0}  ,
  \end{array} \\[3mm]
  \label{eq:107}
  \fl
  &1^-\mathrm{-DSW:}\qquad
  \begin{array}{ll}
    \mbox{linear degeneracy} & {\displaystyle \frac{\partial
        v_-}{\partial \rho_2} = 
    0 \quad \mathrm{or} \quad \frac{\partial s_-}{\partial \rho_1} =
    0} , \\
    \mbox{gradient catastrophe} & {\displaystyle \frac{\partial
        v_-}{\partial \rho_1} = 
    0 \quad \mathrm{or} \quad \frac{\partial s_-}{\partial \rho_2} =
    0}  .
  \end{array}
\end{eqnarray}
\endnumparts
The negation of these breakdown criteria are further necessary
admissibility criteria, additional to (\ref{eq:283})--(\ref{eq:68}),
for the validity of the simple wave DSW construction.

\section{Weak DSWs}
\label{sec:weak-dsws}

The jump conditions for an admissible 1-DSW presented in section
\ref{sec:normal-dsws} apply generally to dispersive Eulerian fluids
satisfying hypotheses A1-A5.  They can be determined explicitly in the
case of weak DSWs.  An asymptotic analysis of the jump conditions is
presented below assuming a weak 1-DSW corresponding to a small jump in
density
\begin{equation*}
    \rho_2 = \rho_1 + \Delta, \qquad |\Delta |  \ll 1 .
\end{equation*}
Two approaches are taken.  First, asymptotics of the Whitham-El simple
wave DSW closure theory are applied and second, direct KdV asymptotics
of the dispersive Euler equations are used.

\subsection{Linear Wave Edge}
\label{sec:linear-wave-edge}

Expanding the linear wave edge speed (\ref{eq:75}) yields
\begin{equation}
  \label{eq:157}
  v_j(\rho_1,\rho_1 + \Delta) \sim \lim_{\rho_2 \to \rho_1}
  v_j(\rho_1,\rho_2) + \pd{}{\rho_2} v_j(\rho_1,\rho_2) \Delta .
\end{equation}
Using the long wave asymptotics of the dispersion relation
(\ref{eq:80}) and the initial condition for the integral curve
(\ref{eq:73}), the first term is
\begin{equation*}
  \lim_{\rho_2 \to \rho_1} v_j(\rho_1,\rho_2) = u_1 - \lim_{k \to 0} \,
  \omega_{0_k} = u_1 - c_1.
\end{equation*}
The derivative term in (\ref{eq:157}) for the case $j = 1$ is
evaluated using (\ref{eq:197})--(\ref{eq:75})
\begin{eqnarray}
  \label{eq:159}
  \eqalign{
  \lim_{\rho_2 \to \rho_1}
  \pd{}{\rho_2} v_1(\rho_1,\rho_2) &=  \lim_{\rho_2 \to \rho_1}
  - \omega_{0_{kk}} \pd{k}{\rho_2} = \lim_{\rho_2 \to \rho_1}
  \omega_{0_{kk}} \frac{\rmd k}{\rmd \rho_1} \\
  &= \lim_{k\to 0} \omega_{0_{kk}} \frac{c_1 k/\rho_1 + \omega_{0_{\rhobar}}}{c_1 -
    \omega_{0_k}} \\
  &= -2 \left ( \frac{c_1}{\rho_1} + c_1' \right ) .}
\end{eqnarray}
The second equality in (\ref{eq:159}) involves differentiation with respect
to the initial ``time'' $\rho_2$ which, due to uniqueness of solutions to the
initial value problem, satisfies
\begin{equation}
  \label{eq:148}
  \pd{k}{\rho_2}(\rho_1;\rho_2) = - \frac{\rmd
    k}{\rmd \rho_1}(\rho_1;\rho_2) .
\end{equation}
The last equality in (\ref{eq:159}) follows from the weak dispersion
asymptotics (\ref{eq:80}).  A similar computation for the $j = 2$ case
gives
\begin{eqnarray*}
  \lim_{\rho_2 \to \rho_1} \pd{}{\rho_2} v_2(\rho_1,\rho_2) &=
  \lim_{\rho_2 \to \rho_1} \ubar^{\, \prime} (\rho_2) -
  \omega_{0_{k\rhobar}} - \omega_{0_{kk}} \frac{\rmd k}{\rmd \rhobar}
  \\
  &=  -\frac{c_1}{\rho_1} - \left ( \lim_{k \to 0} \omega_{0_{k\rhobar}} +
    \omega_{0_{kk}} \frac{c_1 k/\rho_1 + \omega_{0_{\rhobar}}}{c_1 -
      \omega_{0_k}} \right ) \\
  &= \frac{c_1}{\rho_1} + c_1' .
\end{eqnarray*}
Combining this result with the other speed calculation gives
\begin{equation}
  \fl
  \label{eq:156}
  v_j(\rho_1,\rho_1+ \Delta)  \sim u_1 - c_1 + (-1)^j (3-j)
  \left ( \frac{c_1}{\rho_1} + c_1' 
  \right ) \Delta, \qquad |\Delta | \ll 1 .
\end{equation}

The corresponding wavenumber at the linear wave edge can also be
determined perturbatively.  Note that a Taylor expansion of $k_1 =
k(\rho_1;\rho_1 + \Delta)$ for small $\Delta$ is not valid
because $k(\rho;\rho_1)$ is not analytic in a neighborhood of
$\rho_1$, exhibiting a square root branch point.  However, $k_1^2$ is
analytic, so that upon Taylor expansion, the use of (\ref{eq:148}) and
(\ref{eq:80}) yield
\begin{equation*}
  k_j \sim \frac{2}{3 | \mu|} \left ( \frac{c_1}{\rho_1} + c_1' \right
  ) \sqrt{|\Delta|} , \qquad j = 1,2, \qquad |\Delta | \ll 1 ,
\end{equation*}
the wavenumber of the linear wavepacket at a weak $1^\pm$-DSW's linear
wave edge.  Note that the wavenumber is independent of the sign of
dispersion.

\subsection{Soliton Edge}
\label{sec:jump-conditions}

The soliton edge speed is expanded for a small density jump as
\begin{equation*}
  s_j(\rho_1,\rho_1+\Delta) = \lim_{\rho_2 \to \rho_1}
  s_j(\rho_1,\rho_2) + \pd{}{\rho_2} 
  s_j(\rho_1,\rho_2) \Delta + \cdots .
\end{equation*}
Using the expansion (\ref{eq:80}), the definition (\ref{eq:26}), the
expression (\ref{eq:25}), and the initial condition (\ref{eq:53})
gives
\begin{equation*}
  \lim_{\rho_2 \to \rho_1} s_j(\rho_1,\rho_2) = u_1 - \lim_{\tilde{k} \to
    0} \frac{\tilde{\omega}_0(\tilde{k},\rho_1)}{\tilde{k}} = u_1 -c_1 .
\end{equation*}

To compute the limit $\pd{}{\rho_2} s_j(\rho_1,\rho_1)$ necessitates
different considerations for each $j$.  When $j = 1$,
(\ref{eq:25}) gives
\begin{eqnarray*}
    \lim_{\rho_2 \to \rho_1} \pd{}{\rho_2}  s_1(\rho_1,\rho_2) &= 
    \lim_{\rho_2 \to \rho_1} -
    \frac{\omegat_{0_{\tilde{k}}}\tilde{k} 
      - \tilde{\omega}_0}{\tilde{k}^2 }
  \pd{\tilde{k}}{\rho_2} \\
  &= \lim_{\rho_2 \to \rho_1} 
    \frac{\omegat_{0_{\tilde{k}}}\tilde{k} 
      - \tilde{\omega}_0}{\tilde{k}^2 }
  \frac{\rmd \tilde{k}}{\rmd \rho_1} \\
  &=
  \lim_{\tilde{k} \to 0} \frac{ (\tilde{\omega}_{0_{\tilde{k}}} \kt
    - \tilde{\omega}_0)( c_1 \kt/\rho_1 + \omegat_{0_{\rhobar}}
    )}{\kt^2( c_1 - \omegat_{0_{\kt}})} \\
  &= - \frac{2}{3} \left ( \frac{c_1}{\rho_1} + c_1' \right ) .
\end{eqnarray*}
When $j = 2$, the limit is similarly computed as
\begin{eqnarray*}
  \fl
  \lim_{\rho_2 \to \rho_1} \pd{}{\rho_2} s_2(\rho_1,\rho_2)
  &= \lim_{\rho_2 \to \rho_1} u'(\rho_1) -
  \frac{\omegat_{0_{\rhobar}}}{\kt} - \frac{\omegat_{0_{\kt}} \kt -
    \omegat_0}{\kt^2} \frac{d\kt}{d \rhobar} \\
  &= \lim_{\tilde{k} \to 0} -\frac{c_1}{\rho_1} - 
  \frac{\omegat_{0_{\rhobar}}}{\kt} - \frac{(\omegat_{0_{\kt}} \kt -
    \omegat_0)(c_1 \kt/\rho_1 + \omegat_{0_{\rhobar}})}{\kt^2(c_1 -
    \omegat_{0_{\kt}}) } \\
    &= - \frac{1}{3} \left ( \frac{c_1}{\rho_1} + c_1' \right ) .
\end{eqnarray*}
Combining these results gives the asymptotic soliton edge speed
\begin{equation}
  \label{eq:138}
  s_j(\rho_1,\rho_1 + \Delta) \sim u_1 - c_1 - \frac{3-j}{3} \left (
    \frac{c_1}{\rho_1} + c_1' \right ) \Delta, \qquad |\Delta
  | \ll 1 . 
\end{equation}

\subsection{Admissibility:  Positive and Negative Dispersion}
\label{sec:admissibility}

By insertion of the DSW speeds (\ref{eq:156}) and (\ref{eq:138}) into
the general admissibility criteria, it is found that the $1^+$-DSW
criteria (\ref{eq:283})--(\ref{eq:68}) are satisfied if and only if
$\Delta > 0$ and $\mathrm{sgn}\, \omega_{kk} > 0$.  Similarly,
the $1^-$-DSW criteria (\ref{eq:281})--(\ref{eq:282}) hold if and only
if $\Delta > 0$ and $\mathrm{sgn}\, \omega_{kk} < 0$.  Hence, the
notation $1^\pm$-DSW associated with the dispersion sign is justified
for weak DSWs.

In the notation of (\ref{eq:103}), the weak $1^\pm$-DSW speeds are
\numparts
\begin{eqnarray}
  \fl
  \label{eq:203}
  s_\pm^{(1)}(\rho_1,\rho_1 + \Delta) \sim u_1 - c_1 -
  \frac{3 \mp 1}{6} \left ( \frac{c_1}{\rho_1} + c_1' \right ) \Delta, \\
  \fl
  \label{eq:204}
  v_\pm^{(1)}(\rho_1,\rho_1+ \Delta) \sim u_1 - c_1 - \frac{1 \pm
    3}{2}
  \left ( \frac{c_1}{\rho_1} + c_1' \right ) \Delta, 
  \qquad 0 < \Delta \ll 1 ,
\end{eqnarray}
\endnumparts
where the superscript denotes the association with a 1-DSW.  Notably,
the DSW speeds (\ref{eq:203}) and (\ref{eq:204}) differ from the
dissipatively regularized shock speed (\ref{eq:98}) only in the
numerical coefficient of the $(c_1/\rho_1 + c_1')\delta\rho$ term.  A
similar analysis shows that the 2-DSW locus (\ref{eq:136}) requires a
\emph{negative} jump in density and yields the speeds
\begin{eqnarray*}
  \fl
  s_\pm^{(2)}(\rho_2 + \Delta ,\rho_2) \sim u_2 + c_2 + \frac{3 \mp
    1}{6} \left ( \frac{c_2}{\rho_2} + c_2' \right ) \Delta, \\ 
  \fl
  v_\pm^{(2)}(\rho_2 + \Delta,\rho_2) \sim u_2 + c_2 + \frac{1 \pm
    3}{2} \left ( \frac{c_2}{\rho_2} + c_2' \right ) \Delta,
  \qquad 
  0 < \Delta \ll 1 .
\end{eqnarray*}

\subsection{Stationary Soliton Edge}
\label{sec:stat-solit-edge-1}

Choosing the reference frame moving with the $1^\pm$-DSW soliton edge
so that $s_\pm^{(1)} = 0$ results in the relations
\begin{eqnarray*}
  \Delta_\pm &\sim \frac{2 \rho_1 c_1}{(1 \mp 1/3) (c_1 + \rho_1
    c_1')} (M_1 - 1), \\
  M_{2,\pm} &\sim 1 - 2^{\pm 1} (M_1 - 1), \qquad 0 < M_1 - 1 \ll 1 ,
\end{eqnarray*}
which differ from their classical counterparts (\ref{eq:127}) only by
a numerical coefficient.  Upstream supersonic flow through a weak,
admissible DSW results in downstream subsonic flow as in the classical
case.

\subsection{KdV DSWs}
\label{sec:kdv-dsws}

An alternative method to derive weak DSW properties is to consider the
weakly nonlinear behavior of the dispersive Euler equations directly.
Inserting the multiple scales expansion
\begin{eqnarray*}
  \rho &= \rho_1 + \Delta \rho^{(1)}(\xi,T) + \Delta^2
  \rho^{(2)}(\xi,T) 
  + \cdots, \\
  u &= u_1 - \Delta u^{(1)}(\xi,T) + \Delta^2
  u^{(2)}(\xi,T) + \cdots, 
  \\
  \xi &= \Delta^{1/2} [x - (u_1 - c_1) t], \qquad T = \Delta^{3/2} t,
  \qquad 0 < \Delta \ll 1 , 
\end{eqnarray*}
into (\ref{eq:1}), equating like powers of $\Delta$ to
$\mathcal{O}(\Delta^{5/2})$, and recalling the assumed small
wavenumber behavior of the dispersion relation (\ref{eq:80}) yields
the KdV equation
\begin{equation}
  \label{eq:209}
  u^{(1)}_T - \left ( 1 + \frac{\rho_1 c_1'}{c_1} \right ) u^{(1)}
  u^{(1)}_\xi + \mu u^{(1)}_{\xi\xi\xi} = 0 , \qquad \rho^{(1)} =
  \frac{\rho_1}{c_1} u^{(1)} .
\end{equation}
The initial data (\ref{eq:77}) along the 1-DSW locus (\ref{eq:20})
leads to the identification
\begin{equation}
  \label{eq:229}
  u^{(1)}(\xi,0) = 
  \left \{ \begin{array}{ll}
    0  & \xi < 0 \\
    c_1/\rho_1 & \xi > 0 .
  \end{array} \right .
\end{equation}
The large $T$ behavior of $u^{(1)}$ satisfying (\ref{eq:209}) for the
initial data~(\ref{eq:229}) results in a DSW whose structure and edge
speeds depend on $\mathrm{sgn}\, \mu$.  For negative dispersion, $\mu
< 0$, the DSW is oriented such that the leading, leftmost edge is
characterized by a positive, bright soliton.  The positive dispersion
case is equivalent to the negative dispersion case by the
transformations $x \to - x$, $t \to - t$, and $u^{(1)} \to - u^{(1)}$.
Therefore, the leading, leftmost edge is the linear wave edge and the
trailing edge is characterized by a negative, dark soliton.  Scaling
the KdV DSW speeds back to the $(x,t)$ variables results precisely in
the admissible, approximate weak $1^\pm$-DSW speeds (\ref{eq:203}) and
(\ref{eq:204}).  The KdV DSW provides additional information, the
approximate amplitude of the soliton edge 
\begin{eqnarray*}
  1^+\mathrm{-DSW:} \qquad \rho(s_+ t, t) \sim
  \rho_1 \left ( 1 - \frac{\rho_1}{c_1^2} \Delta \right ), \\
  1^-\mathrm{-DSW:} \qquad \rho(s_- t,t) \sim
  \rho_1 \left ( 1 + 2 \frac{\rho_1}{c_1^2} \Delta \right ) .
\end{eqnarray*}

\begin{figure}
  \centering
  \includegraphics[scale=1]{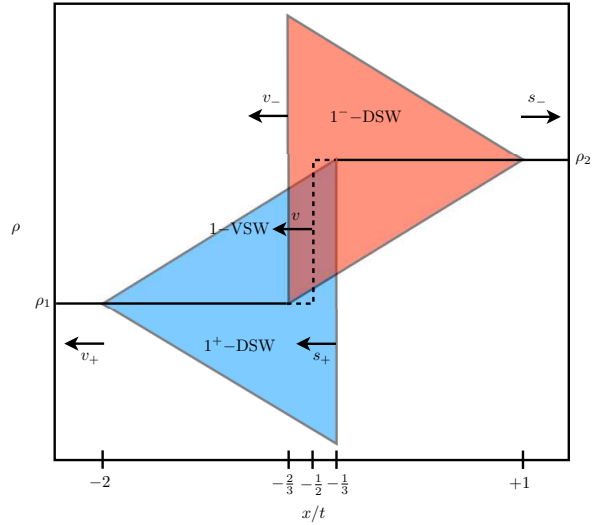}
  \caption{Universal properties of weak $1^\pm$-DSWs with a weak 1-VSW
    (viscous shock wave).  DSWs are represented by their envelopes and
    edge speeds.  Speeds are in units of $(c_1/\rho_1 + c_1') \Delta$
    and $u_1 = c_1$ for simplicity.}
  \label{fig:weak_dsw_schematic}
\end{figure}

\subsection{Discussion}
\label{sec:discussion}

The analysis of this section has yielded the behavior of weak
Eulerian DSWs in the context of assumptions A1-A5 and the
admissibility criteria (\ref{eq:283})--(\ref{eq:282}).  For classical,
weak 1-shocks (2-shocks), the dispersionless Riemann invariant $r_2$
($r_1$) is constant across the shock to third order in the jump height
$\Delta$ \cite{whitham_linear_1974}.  Recalling that the DSW loci
(\ref{eq:20}), (\ref{eq:136}) correspond to constancy of a Riemann
invariant (simple wave condition), the classical Hugoniot loci and the
DSW loci for weak shocks are the same to $\mathcal{O}(\Delta^3)$.
However, the shock speeds differ at $\mathcal{O}(\Delta)$.  The
universal properties of weak shocks regularized by positive/negative
dispersion and dissipation are depicted in figure
\ref{fig:weak_dsw_schematic}.  The jump height $\Delta$, upstream
density $\rho_1$, and pressure law $P(\rho)$ impart only a uniform
scaling of the shock speeds by $(c_1/\rho_1 + c_1') \Delta$ and a
relative scaling of the $1^\pm$-DSW soliton amplitudes by $\rho_1
\Delta/c_1^2$.  All shock speeds differ, showcasing the distinguishing
properties of each regularization type.  $1^-$-DSWs exhibit
backpropagation whereas $1^+$-DSWs and $1$-VSWs (viscous shocks) do
not.  The prominent soliton edge of a $1^-$-DSW ($1^+$-DSW) propagates
faster (slower) than a classical shock.  By continuity and the
discussion of admissibility, it is expected that moderate amplitude
DSWs for Eulerian fluids with a fixed sign of the dispersion exhibit a
structure similar to that pictured in figure
\ref{fig:pos_neg_dispersion}.  This is indeed the case for the example
fluids considered in this work, see section \ref{sec:applications}.

\section{Dispersive Breaking Time}
\label{sec:dsw-breaking-time}

In the small dispersion regime, the hydrodynamic dispersionless system
(\ref{eq:196}) asymptotically describes the evolution of smooth
initial data until breaking occurs.  One can therefore use the
breaking time estimates from section \ref{sec:breaking-time} to estimate
the time at which dispersive terms become important.  This result was
applied to the NLS equation with $c(\rho) = \rho^{1/2}$ in
\cite{forest_onset_1998-1} to estimate the onset of oscillations in
fiber optic pulse propagation.  Here, the breaking time estimates are
applied to polytropic gases with sound speeds $c(\rho) = p^{1/2}
\rho^{p/2}$, $0 < p < 2$ (e.g., gNLS with power law nonlinearity and
gSerre).  Generalizations using Lax's theory developed in section
\ref{sec:breaking-time} are straightforward.

The flow considered is a slowly varying Gaussian on a quiescent
background
\begin{equation}
  \label{eq:2}
  u(x,0) = 0, \qquad \rho(x,0) = 1 + \exp[-(\eps x)^2] , \qquad 0 <
  \eps \ll 1. 
\end{equation}
Slowly varying initial data ensures the applicability of the
dispersionless system (\ref{eq:196}) up to breaking when $t =
\mathcal{O}(1/\eps)$.  This choice of initial data has been used
in photonic DSW experiments \cite{barsi_spatially_2012}.

\begin{figure}
  \centering
  \includegraphics[scale=1]{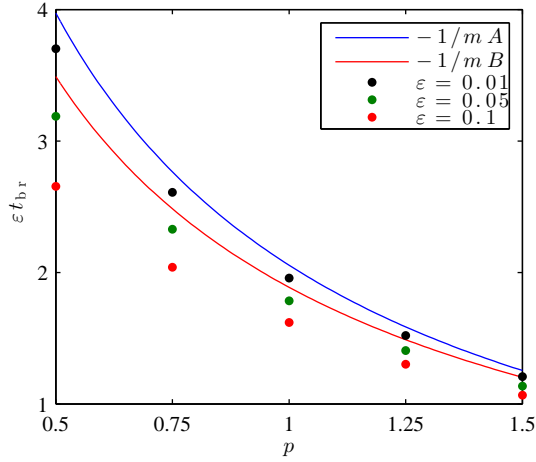}
  \caption{Breaking time bounds (\full) and numerically computed
    breaking times (\fullcircle) for gNLS with power law nonlinearity
    $f(\rho) = \rho^p$ and slowly varying Gaussian initial data
    (\ref{eq:2}).}
  \label{fig:breaking_times}
\end{figure}

Figure \ref{fig:breaking_times} shows the results for the gNLS
equation (\ref{eq:83}) with power law nonlinearity $f(\rho) = \rho^p$.
The solid curves (\full) correspond to the upper and lower bounds on
the dispersionless breaking time estimates (\ref{eq:319}) and
(\fullcircle) correspond to numerically computed breaking times for
several choices of $p$ and $\eps$.  The breaking time from simulations
is defined to be the time at which the density first develops one full
oscillation in the breaking region.  For $|\eps| \ll 1$, the bounds
(\ref{eq:319}) accurately estimate the breaking time across a range of
nonlinearities $p$.  Note that, for these dispersive Euler models and
class of initial data, $\eps \lessapprox 0.01$ is required to obtain
agreement with the dispersionless estimates.

\section{Large Amplitude gNLS DSWs}
\label{sec:applications}

The general Whitham-El simple wave DSW theory is now implemented for
the gNLS equation.

To simplify the presentation, and without loss of generality, the
independent and dependent variables in (\ref{eq:67}) will be scaled so
that the initial jump in density (\ref{eq:77}) is positive, from unit
density
\begin{equation*}
  \rho_1 = 1, \qquad \rho_2 = \Delta > 1 ,
\end{equation*}
so that $1$-DSWs will be considered.  Then, according to the 1-DSW
locus (\ref{eq:20}) of admissible states, the jump in velocity
satisfies
\begin{equation*}
  u_2 = u_1 - \int_1^\Delta \left ( \frac{f'(\rho)}{\rho} \right )^{1/2} \,
  \rmd \rho . 
\end{equation*}

\subsection{General Properties}
\label{sec:general-properties}

The gNLS equation exhibits positive dispersion so that DSWs are of the
$1^+$ variety.  Substituting the expressions (\ref{eq:89}) and
(\ref{eq:152}) into the ODEs (\ref{eq:69}) and (\ref{eq:3}) results in
the initial value problems
\begin{eqnarray}
  \label{eq:166}
  \frac{\rmd k}{\rmd \rhobar} = -k \frac{\displaystyle
    \frac{1}{\rhobar} \left[ 1 +
     \frac{k^2}{4 \rhobar f'(\rhobar)} \right ]^{1/2} +
    \frac{1}{2\rhobar} + \frac{f''(\rhobar)}{2
      f'(\rhobar)}}{\displaystyle 1 +
    \frac{k^2}{2 \rhobar
      f'(\rhobar)} -  \left [1 + \frac{k^2}{4 \rhobar
        f'(\rhobar)} \right ]^{1/2}}, \qquad
  k(\Delta) = 0 .
\end{eqnarray}
for the determination of the linear wave edge speed and
\begin{eqnarray}
  \label{eq:167}
  \frac{\rmd \kt}{\rmd \rhobar} = -\kt \frac{\displaystyle
    \frac{1}{\rhobar} \left [  1 -
      \frac{\kt^2}{4 \rhobar f'(\rhobar)} \right ]^{1/2} +
    \frac{1}{2\rhobar} +
    \frac{f''(\rhobar)}{2 f'(\rhobar)}}{\displaystyle 1 -
    \frac{\kt^2}{2 \rhobar 
      f'(\rhobar)} -  \left [ 1 - \frac{\kt^2}{4 \rhobar
        f'(\rhobar)} \right ]^{1/2} }, \qquad \tilde{k}(1) = 0 ,
\end{eqnarray}
for the soliton edge speed.  In (\ref{eq:167}), it is possible for the
quantity within the square roots to pass through zero.  When this
occurs, an appropriate branch of the dispersion relation should be
used so that the conjugate wavenumber remains real valued.  

Recalling remark \ref{sec:soliton-edge-1} in section
\ref{sec:soliton-edge}, the transformation
\begin{equation*}
  \alpha(\rhobar) = \frac{\omega_0(k,\rhobar)}{c(\rhobar) k} = \left [ 1
    + \frac{k^2}{4 \rhobar f'(\rhobar)}
  \right ]^{1/2},
\end{equation*}
simplifies (\ref{eq:166}) to the ODE
\begin{equation}
  \label{eq:170}
  \frac{\rmd \alpha}{\rmd \rhobar} = -\frac{1}{2}(1 + \alpha) \left [ 
    \frac{1}{\rhobar} + \frac{(2 \alpha - 1) f''(\rhobar)}{(2
      \alpha + 1) f'(\rhobar)} \right ] ,
\end{equation}
with initial condition
\begin{equation}
  \label{eq:163}
  \alpha(\Delta) = 1 .
\end{equation}
The analogous transformation for the conjugate variables
\begin{equation}
  \label{eq:211}
  \alphat(\rhobar) = \frac{\omegat_0(\kt,\rhobar)}{c(\rhobar) \kt} =
  \left [ 1 - \frac{\kt^2}{4
      c(\rhobar)^2} \right ]^{1/2} ,
\end{equation}
transforms (\ref{eq:167}) to the same equation (\ref{eq:170}) with
$\alpha \to \alphat$ and the initial condition
\begin{equation}
  \label{eq:171}
  \tilde{\alpha}(1) = 1 .
\end{equation}

Upon solving the initial value problems for $\alpha$ and $\alphat$,
the linear wave and soliton edge speeds are found from (\ref{eq:75})
and (\ref{eq:25}), respectively, which take the form
\numparts
\begin{eqnarray}
  \label{eq:180}
  v_+ =
  u_1 + \frac{1 - 2 \alpha(1)^2}{\alpha(1)} \sqrt{f'(1)}, \\
  \label{eq:181}
  s_+ = u_1 - 
  \tilde{\alpha}(\Delta) \sqrt{\Delta f'(\Delta)} - \int_{1}^{\Delta}
  \left [ \frac{f'(\rho)}{\rho} \right 
  ]^{1/2} \rmd\rho ,
\end{eqnarray}
\endnumparts
in the $\alpha$, $\alphat$ variables.  The weak DSW results
(\ref{eq:156}) and (\ref{eq:138}) give the approximations
\numparts
\begin{eqnarray}
  \fl
  \label{eq:124}
  v_+ \sim u_1 - \sqrt{f'(1)} \left \{ 1 + \left [ 3 + \frac{f''(1)}{f'(1)}
    \right ] (\Delta - 1) \right \} , \\
  \label{eq:134}
  \fl
  s_+ \sim u_1 -  \sqrt{f'(1)} \left \{ 1 + \frac{1}{6} \left [ 3 +
      \frac{f''(1)}{f'(1)} 
    \right ] (\Delta - 1) \right \} , \qquad 0 < \Delta - 1 \ll 1.
\end{eqnarray}
\endnumparts

Equating the soliton speed in (\ref{eq:181}) to the soliton/amplitude
speed relation in (\ref{eq:93}) gives an implicit relation for the
dark soliton minimum $\rhomin$ in terms of the background density
$\Delta$
\begin{eqnarray}
  \label{eq:187}
  \tilde{\alpha}(\Delta)^2 &\Delta f'(\Delta) = 
  \frac{2 \rhomin}{\Delta - \rhomin} \left | f(\Delta) -
    \frac{1}{\Delta - \rhomin} \int_{\rhomin}^{\Delta} f(\rho) \rmd\rho
  \right | .
\end{eqnarray}

A direct computation shows that neither linear degeneracy
(\ref{eq:137}) nor a sign of dispersion change (\ref{eq:200}) occurs
at the linear wave edge.  However, at the soliton edge, there are
several distinguished values of $\tilde{\alpha}(\Delta)$ with physical
ramifications.  The modulation theory breaks down due to singular
derivative formation in (\ref{eq:170}) when
\begin{equation*}
  \tilde{\alpha}(\Delta = \Deltas) = -\frac{1}{2} .
\end{equation*}
From the denominator in (\ref{eq:3}), singularity formation occurs
precisely when $\omegat_{0_{\kt}} = c(\Delta)$.  A direct computation
shows that $\omegat_0$ is a concave function of $\kt$, which implies
$\omegat_0/\kt > \omegat_{0_{\kt}} = c(\Delta)$ so that singularity
formation coincides with the violation of the admissibility criterion
(\ref{eq:270}).

From the initial condition (\ref{eq:171}) and the ODE (\ref{eq:170}),
$\alphat(\Delta)$ decreases from 1 for increasing $\Delta$
sufficiently close to 1.  The value of $\tilde{\alpha}$ at which its
derivative is zero, from (\ref{eq:170}), is
\begin{equation*}
  \tilde{\alpha}_\mathrm{min}(\Delta) = \frac{\Delta f''(\Delta) -
    f'(\Delta)}{2[\Delta f''(\Delta) + f'(\Delta)]} .
\end{equation*}
So long as
\begin{equation*}
  \tilde{\alpha}(\Delta) > \max \left [ \tilde{\alpha}_\mathrm{min}(\Delta) ,
    -\frac{1}{2} 
  \right ] ,
\end{equation*}
the right hand side of (\ref{eq:170}) is strictly negative
and finite.  Since $\alpha$ satisfies the same ODE as $\alphat$, if
$\tilde{\alpha}_\mathrm{min} < 1$, $\alpha(1)$ is an \emph{increasing}
function for all $\Delta$ and $\tilde{\alpha}(\Delta)$ is a
\emph{decreasing} function of $\Delta$ until $\tilde{\alpha}(\Delta) =
\max(\tilde{\alpha}_\mathrm{min}(\Delta), -1/2)$.

As $\Delta$ is increased from $1$, cavitation (a point of zero density
or \emph{vacuum point}) first occurs when $\rhomin = 0$ corresponding
to a black soliton at the trailing edge, moving with the background
flow speed $\ubar(\Delta)$.  Since necessarily $\Delta > \rhomin$,
(\ref{eq:187}) implies that a vacuum point first occurs when
\begin{equation}
  \label{eq:188}
  \tilde{\alpha}(\Delta = \Deltav) = 0 .
\end{equation}
For larger jumps $\Delta > \Deltav$, the vacuum point is expected to
develop in the interior of the DSW
\cite{el_decay_1995,el_theory_2007}.  According to the transformation
(\ref{eq:211}), when $\alphat$ crosses zero, the branch of the
conjugate dispersion changes sign.

Using Galilean invariance, it is convenient to consider the reference
frame moving with the soliton edge so that $s_+ = 0$.  In such a
frame, given the upstream supersonic flow velocity $u_1 > 1$, the
density jump $\Delta$ is determined from (\ref{eq:181}) and the
downstream flow velocity satisfies $u_2 = \alphat(\Delta) \sqrt{
  \Delta f'(\Delta)}$.  Then, for $\Delta < \Deltav$, $\alphat(\Delta)
> 0$ so that $u_2 > 0$.  But, when $\Delta > \Deltav$, cavitation
occurs and $\alphat(\Delta)$ changes sign causing $u_2 < 0$.
Counterintuitively, the dispersive fluid flows \emph{into} the DSW
from both sides upon the generation of a vacuum point.  This behavior
has been observed in NLS \cite{gurevich_dissipationless_1987} and
photorefractive gNLS \cite{el_theory_2007}.

At the soliton edge, linear degeneracy (\ref{eq:100}) occurs when
\begin{equation}
  \label{eq:179}
  \omegat_{0_{\rhobar}} + \frac{c \kt}{\rhobar} 
  = c \kt \left [ \frac{f'(\Delta) + \Delta f''(\Delta)}{2
      \Delta f'(\Delta) \alphat} +
    \frac{1}{\Delta} \right ]  = 0 . 
\end{equation}
The only way for this to occur is for $\omegat_0$ to change to another
branch of the conjugate dispersion relation, i.e., for $\alphat$ to
pass through 0.  Then from (\ref{eq:179}), the value of $\alphat$ at
which linear degeneracy occurs is
\begin{equation}
    \label{eq:227}
    \tilde{\alpha}_\mathrm{l}(\Delta) = 
    -\frac{1}{2}\left [ 1 + \frac{\Delta f''(\Delta)}{
        f'(\Delta)} \right ] .
\end{equation}
According to the assumptions on $f$ (\ref{eq:139}),
$\alphat_\mathrm{l} < 0$ as required.  Since the linear degeneracy
condition (\ref{eq:179}) amounts to $\rmd s_+/\rmd \Delta = 0$ (recall
(\ref{eq:106})), the distinguished value $\alphat_\mathrm{l}$
coincides with an extremum of the soliton edge speed as the jump
height $\Delta$ is varied.

A direct computation verifies that the zero dispersion criterion
(\ref{eq:302}) does not occur.  The admissibility criteria
(\ref{eq:283}) and (\ref{eq:284}) correspond to $\alphat(\Delta) < 1$
and $\alpha(1) < 1$ which are true generically.  In summary, so long
as $\alphat_\mathrm{min} < -1/2$, $\alphat(\Delta)$ is a decreasing
function that can attain the distinguished values $\alphat = 0$ when
$\Delta = \Delta_\mathrm{v}$ corresponding to a vacuum point or
cavitation, $\alphat = \alphat_\mathrm{l}$ when $\Delta =
\Delta_\mathrm{l}$ coinciding with linear degeneracy and the breakdown
of the simple wave criterion, and $\alphat = -1/2$ when $\Delta =
\Delta_\mathrm{s}$ leading to singularity formation.  For $\Delta <
\Delta_\mathrm{l}$, the only admissibility criterion left to verify is
the leading and trailing edge ordering (\ref{eq:68}).

The DSW regularization is completed upon solving the ODE
(\ref{eq:170}).  This equation is separable for nonlinearity
satisfying $f'(\rho) \pm \rho f''(\rho) = 0$.  Recalling the
admissible nonlinearity (\ref{eq:139}), the cases of interest are
$f(\rho) = \kappa \rho^p$, $p > 0$, corresponding to polytropic
superfluids.  This class of nonlinearity is now considered.

\subsection{Power Law Nonlinearity}
\label{sec:polytropic-gas-frho}

With the choice (\ref{eq:164}) of power law nonlinearity, the speed of
sound is $c(\rho) = p^{1/2} \rho^{p/2}$.  Equation (\ref{eq:170})
becomes the separable equation
\begin{equation}
  \label{eq:212}
  \frac{-2(1 + 2 \alpha)}{(1+ \alpha)(2 - p + 2p \alpha)} \rmd \alpha =
  \frac{\rmd \rhobar}{\rhobar} .
\end{equation}
Using the initial condition (\ref{eq:163}), (\ref{eq:212}) is
integrated to
\numparts
\begin{eqnarray}
  \fl
  \label{eq:172}
  \frac{1 + \alpha(1)}{2} \left [\frac{2 - p + 2p \alpha(1)}{2 + p} \right
  ]^{2(p-1)/p} =  
  \Delta^{(3p - 2)/2}, \qquad p \ne \frac{2}{3} ,\\
  \fl
  \label{eq:173}
  2 \ln \left [ \frac{1 + \alpha(1)}{2} \right ] + \frac{1}{1+\alpha(1)} -
  \frac{1}{2} = \frac{2}{3} \ln \Delta  ,
  \qquad p = \frac{2}{3} .
\end{eqnarray}
\endnumparts
and for (\ref{eq:171}), (\ref{eq:212}) is integrated to
\numparts
\begin{eqnarray}
  \fl
  \label{eq:153}
  \frac{1 + \tilde{\alpha}(\Delta)}{2} \left [\frac{2 - p + 2p
      \tilde{\alpha}(\Delta)}{2 + p} \right ]^{2(p-1)/p} =
  \Delta^{(2- 3p)/2}, \qquad p \ne \frac{2}{3} ,\\
  \fl
  \label{eq:155}
  2 \ln \left [ \frac{1 + \tilde{\alpha}(\Delta)}{2} \right ] +
  \frac{1}{1+\tilde{\alpha}(\Delta)} - 
  \frac{1}{2} = -\frac{2}{3} \ln \Delta ,
  \qquad p = \frac{2}{3} ,
\end{eqnarray}
\endnumparts
the integrals existing so long as
\begin{equation*}
  \alpha(1), ~\tilde{\alpha}(\Delta) > \frac{p - 2}{2 p} =
  \tilde{\alpha}_\mathrm{min} . 
\end{equation*}

The linear wave and soliton edge speeds are then determined via
(\ref{eq:180}) and (\ref{eq:181})
\numparts
\begin{eqnarray}
  \label{eq:213}
  \frac{v_+ - u_1}{c_1} = \frac{1 - 2
    \alpha(1)^2}{\alpha(1)} , \\
  \label{eq:218}
  \frac{s_+-u_1}{c_1} =  - \Delta^{p/2}
  \left [ \frac{2}{p} + \tilde{\alpha}(\Delta) \right ] +
  \frac{2}{p} .
\end{eqnarray}
\endnumparts
Equations (\ref{eq:172}) and (\ref{eq:153}) can be solved explicitly
for $\alpha$ in several cases \numparts
\begin{eqnarray}
  \fl
  \label{eq:225}
  p =\frac{1}{2}: \quad \left \{
    \begin{array}{>{\displaystyle}c}
      \alpha(1) = \frac{25}{16} \Delta^{1/4} -\frac{3}{2} +
      \frac{5}{16} \left (25
        \Delta^{1/2} - 16 \Delta^{1/4} \right )^{1/2}, \\
      \tilde{\alpha}(\Delta) = \frac{25}{16} \Delta^{-1/4} -
      \frac{3}{2} + \frac{5}{16} \left ( 25\Delta^{-1/2} - 16
        \Delta^{-1/4} \right )^{1/2} , 
    \end{array} \right . 
  \\
  \fl
  \label{eq:217}
  p = 1: \quad \left \{
    \begin{array}{c}
      \alpha(1) = 2 \Delta^{1/2} - 1, \\
      \tilde{\alpha}(\Delta) = 2  \Delta^{-1/2} - 1,
    \end{array} \right .
  \\
  \fl
  \label{eq:219}
  p = 2: \quad \left \{
    \begin{array}{>{\displaystyle}c}
      \alpha(1) =\frac{1}{2} \left ( \sqrt{1 + 8 \Delta^2} -
      1 \right ), \\
      \tilde{\alpha}(\Delta) =
      \frac{1}{2} \left (\sqrt{1 + 8\Delta^{-2}} - 1 \right ),
    \end{array} \right .
\end{eqnarray}
\endnumparts
providing explicit expressions for the 1-DSW linear wave and soliton
edge speeds
\numparts
\begin{eqnarray}
  \fl
  \label{eq:226}
  p = \frac{1}{2}: \quad 
  \left \{ 
    \begin{array}{>{\displaystyle}l}
      \frac{v_+-u_1}{c_1} =  \frac{16 - \frac{1}{8}
        \left [ 25 \Delta^{1/4} - 24 + 5 \sqrt{ 25 \Delta^{1/2} - 16
            \Delta^{1/4}} \right ]^2}{25 \Delta^{1/4} -
        24 + 5  \sqrt{25 \Delta^{1/2} - 16 \Delta^{1/4}}}, \\[2mm]
      \frac{s_+ - u_1}{c_1} = - \frac{5}{2} \Delta^{1/4} +
      \frac{39}{16} - \frac{5}{16} \sqrt{25 - 16 \Delta^{1/4} } ,
    \end{array} \right .
  \\
  \fl
  \label{eq:175}
  p = 1: \quad
  \left \{ 
    \begin{array}{>{\displaystyle}l}
      \frac{v_+ - u_1}{c_1} =  - \frac{8
          \Delta -8 \Delta^{1/2} + 1}{2 \Delta^{1/2} - 1}, \\[2mm]
      \frac{s_+ - u_1}{c_1} =  - \Delta^{1/2},
    \end{array} \right .
  \\
  \fl
  \label{eq:176}
  p = 2: \quad
  \left \{ 
    \begin{array}{>{\displaystyle}l}
      \frac{v_+-u_1}{c_1} =  - 2 \,
        \frac{4 \Delta^2 -  
          \sqrt{1 + 8 \Delta^2} 
        }{\sqrt{1 + 8 \Delta^2} - 1}, \\[2mm]
        \frac{s_+ - u_1}{c_1} = - \frac{1}{2} \left ( \Delta -
        2 + \sqrt{8 
          + \Delta^2} \right ) .
    \end{array} \right .  
\end{eqnarray}
\endnumparts
The cubic NLS case (\ref{eq:175}) agrees with the original result in
\cite{gurevich_dissipationless_1987}.

The distinguished values of the jump height $\Delta$ predict different
DSW behavior as $p$ varies.  The large $\Delta$ behavior of
(\ref{eq:153}) and the fact that $\tilde{\alpha}(\Delta)$ is a
decreasing function of $\Delta$ proves that
\begin{equation}
  \label{eq:182}
  \tilde{\alpha}(\Delta) \searrow \frac{p - 2}{2 p} > -\frac{1}{2},
  \qquad \Delta \to \infty, \qquad  p > 1 ,
\end{equation}
precluding the possibility of singularity formation when $p > 1$.
Furthermore, when $p \ge 2$, (\ref{eq:182}) prohibits cavitation in
the DSW because $\tilde{\alpha}(\Delta) > 0$.

The cavitation condition (\ref{eq:188}) can be used along with
(\ref{eq:153}) to determine $\Deltav(p)$, the value of $\Delta$
at which a vacuum point is predicted to appear
\numparts
\begin{eqnarray}
  \fl
  \label{eq:189}
  &\Deltav(p) =  2^{2/(3p-2)} \left ( \frac{2-p}{2+p} \right
  )^{4(1-p)/[p(3p - 2)]}, \qquad 0 < p < 2, \qquad p \ne \frac{2}{3} ,
  \\
  \fl
  \label{eq:190}
  &\lim_{p \to 2/3} \Deltav(p) = \frac{8}{\exp(3/4)} \approx 3.78 .
\end{eqnarray}
\endnumparts
For $p$ approaching $2$, the vacuum jump $\Deltav$ increases without
bound.  The limiting behavior for $p \to 0$ is $\Deltav(p) \to
\e^2/2 \approx 3.69$.

The DSW soliton speed exhibits a minimum (maximum in absolute value)
for $0 < p < 1$ corresponding to the onset of linear degeneracy in the
modulation system.  The value of the jump height $\Delta_\mathrm{l}$
at the linear degeneracy point is found by inserting (\ref{eq:227})
into (\ref{eq:153}) and solving for $\Delta$
\numparts
\begin{eqnarray}
  \label{eq:221}
  \fl
  \Delta_\mathrm{l}(p)
  = \left [ \frac{1}{4} (2 - p) ( 1 -
    p)^{2-2/p} \right ]^{2/(2 - 3p)}, \qquad 0 < p < 
  1, \qquad p \ne \frac{2}{3} , \\
  \label{eq:224}
  \fl
  \Delta_\mathrm{l}(2/3) = \frac{27}{\exp(3/2)} \approx 6.02.
\end{eqnarray}
\endnumparts
The $p \to 0$ behavior is $\lim_{p \to 0} \Delta_\mathrm{l}(p) = \e^2/2$,
coincidentally the same limiting value $\Deltav(0)$.  The maximum
value of the trailing edge speed $|s_+ |$ is therefore
\begin{equation*}
  \fl
  \max_{1 < \Delta < \Deltas} | s_+ | = u_1 + c_1 \left | 
    \frac{4 - p^2}{2p} \Delta_\mathrm{l}(p)^{p/2} - \frac{2}{p} \right
  | , \qquad 0 < p < 1 . 
\end{equation*}
For $\Delta > \Delta_\mathrm{l}(p)$, the simple wave construction is
no longer valid.

For $0 < p < 1$, $\alphat(\Delta)$ exhibits singularity formation in
its derivative when $\tilde{\alpha}(\Delta) \to -1/2$ for $\Delta \to
\Deltas(p)$ where
\numparts
\begin{eqnarray}
  \label{eq:183}
  \fl
  \Deltas(p) = 2^{4/[p(3p-2)]} \left ( \frac{1 - p}{2
      + p} \right )^{4(1-p)/[p(3p - 2)]}, \qquad 0 < p < 1, \qquad
  p \ne 2/3, \\
  \fl
  \label{eq:151}
  \Deltas(2/3) = 64 \exp(-9/4) \approx 6.75 ,
\end{eqnarray}
\endnumparts
so that DSWs with $\Delta > \Deltas(p)$ are not admissible.

Based on the foregoing analysis and upon verification of the DSW edge
ordering (\ref{eq:68}), the modulation theory presented here does not
violate any of the admissibility criteria for all $1 < \Delta$ when $p
\ge 1$ and for $1 < \Delta < \Delta_\mathrm{l}(p)$ when $0 < p < 1$.

\subsubsection{Numerical Results}
\label{sec:numerical-results}

\begin{figure*}[!htb]
  \centering 
  \subfigure[{Shock tube problem}\label{fig:shock_tube_density}]{\includegraphics[scale=1]{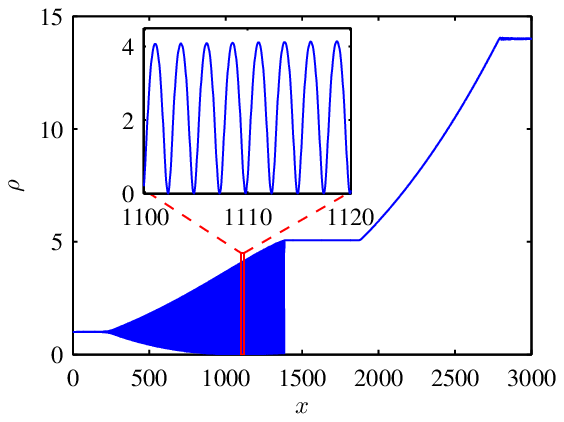}}
  \subfigure[{Shock tube intermediate
    density}\label{fig:shock_tube_intermediate_p}]{\includegraphics[scale=1]{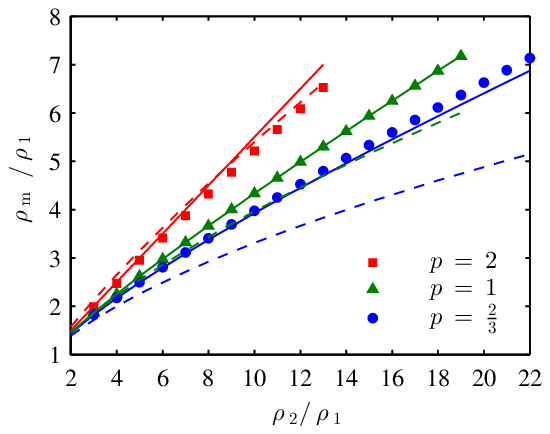}}
  \caption{(a) Density resulting from the numerical solution of the
    shock tube problem for power law gNLS, $p = 2/3$.  The inset
    depicts the slowly varying nature of the DSW.  (b) Predicted
    (\full) and computed ({\color{red} \fullsquare}, {\color{grn}
      \fulltriangle}, {\color{blue} \fullcircle}) intermediate states
    $\rhom$ resulting from the shock tube problem. The prediction for
    a classical shock tube is also given (\dashed).}
  \label{fig:shock_tube}
\end{figure*}

The simple wave DSW construction makes several predictions that can be
tested through long-time numerical simulations of the shock tube
problem of section \ref{sec:disp-riem-probl}.  For the details of the
numerical method, see the appendix.  Recall that the shock tube
problem results in the generation of two waves, a DSW followed by a
rarefaction wave.  Figure \ref{fig:shock_tube_density} depicts the
numerically computed density at $t = 350$ resulting from the initial
jump $(\rho_1,\rho_2) = (1,14)$ at $x = 2000$ for power law gNLS with
$p =2/3$.  The structure of a $1^+$-DSW connected to a 2-rarefaction
via an intermediate, constant state is clear.  Almost imperceptible
modulations of the zoomed-in solution in the inset demonstrate the two
scale nature of dispersive hydrodynamics.  Figure
\ref{fig:shock_tube_intermediate_p} provides a comparison of the
intermediate density $\rhom$ predicted by (\ref{eq:295}) (\full),
numerical computation ({\color{red} \fullsquare}, {\color{grn}
  \fulltriangle}, {\color{blue} \fullcircle}), and the intermediate
density predicted for a classical, polytropic gas, equations
(\ref{eq:297}) (\dashed).  The integrable NLS case $p = 1$ agrees
precisely with the numerical computations but for the non-integrable
cases, eventual deviation is observed for large initial jumps.
Interestingly, the classical shock prediction begins to agree with the
$p = 2$ numerics for sufficiently large jumps.  Similar behavior was
noted for gNLS with photorefractive nonlinearity \cite{el_theory_2007}
and for the Serre equations \cite{el_unsteady_2006}.

\begin{figure*}[!htb]
  \centering
  \subfigure[{Soliton edge speeds}\label{fig:dsw_soliton_speeds_p}]
  {\includegraphics[scale=1]{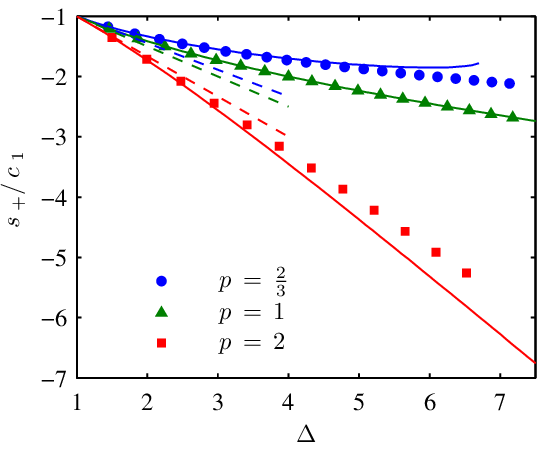}}
  \quad 
  \subfigure[{Linear wave edge speeds}\label{fig:dsw_linear_speeds_p}]
  {\includegraphics[scale=1]{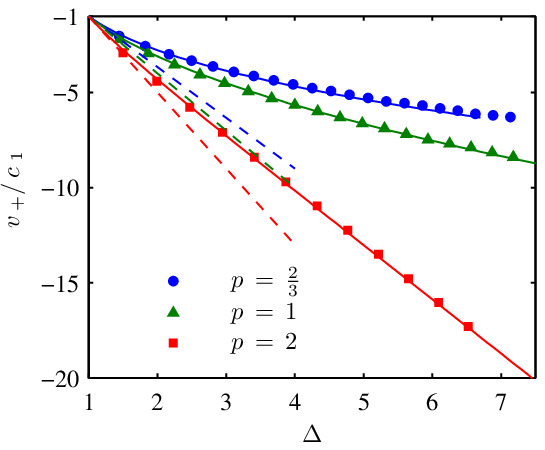}}
  \caption{Power law gNLS 1-DSW speeds for varying density jump
    $\Delta$ and nonlinearity exponent $p$.  Solid curves (\full) are
    from (\ref{eq:218}), (\ref{eq:213}), ({\color{red} \fullsquare},
    {\color{grn} \fulltriangle}, {\color{blue} \fullcircle}) result
    from numerical simulation, and (\dashed) are the weak DSW results
    (\ref{eq:124}), (\ref{eq:134}).} \label{fig:dsw_speeds_p}
\end{figure*}

A comparison of the numerically extracted $1^+$-DSW speeds and simple
DSW theory is shown in figure \ref{fig:dsw_speeds_p}(a,b).  The
speeds, resulting from simulations with different $p$ values, are
normalized by the sound speed $c_1$.  The value of the DSW jump
$\Delta$ corresponding to the numerical simulation results
({\color{red} \fullsquare}, {\color{grn} \fulltriangle}, {\color{blue}
  \fullcircle}) is the one extracted from the computation, i.e.,
$\Delta$ is the numerical value of the density between the DSW and the
rarefaction.  The $\Delta$ value used for the theoretical predictions
(\full, \dashed) is $\rhom$ from the shock tube problem
(\ref{eq:295}).  The computations for the integrable case $p = 1$
agree excellently, even for large jumps.  The solid curves (\full)
show rapid deviation from the weak DSW straight line (\dashed)
predictions, signaling the importance of nonlinearity.  In the weak to
moderate jump regime $\Delta \lesssim 3$, the non-integrable cases
exhibit good agreement with the predicted soliton edge speed but
deviation occurs for large jumps.  Interestingly, the non-integrable
cases admit excellent agreement with the predicted linear wave edge
speed across their entire regions of validity.

\begin{figure}
  \centering
  \includegraphics[scale=1]{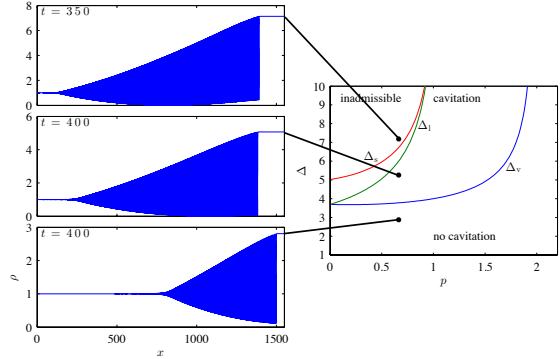}
  \caption{Right: singular $\Delta > \Deltas(p)$, linearly degenerate
    $\Deltas(p) > \Delta > \Delta_\mathrm{l}(p)$, and cavitation
    $\Delta_\mathrm{l}(p) > \Delta > \Deltav(p)$ regions in
    $(\Delta,p)$ phase space for a gNLS simple $1^+$-DSW with power
    law $p$ nonlinearity.  Left: numerically computed $1^+$-DSWs for
    $p = 2/3$.}
  \label{fig:delta_plot}
\end{figure}

Recall that for $p = 2/3$, linear degeneracy sets in for jump heights
above that which corresponds to the minimum of the soliton edge speed
curve (\ref{eq:224}), $\Delta_\mathrm{l} \approx 6$.  However, the
computed $1^+$-DSW shows no distinct change in structure for jumps
below and above this threshold.  This feature is depicted in figure
\ref{fig:delta_plot}.  On the right, regions in $(p,\Delta)$ space
where a vacuum point is predicted to develop, the DSW modulation
theory breaks down due to linear degeneracy, and singular behavior are
shown.  It is clear from this figure that $\Deltav < \Delta_\mathrm{l}
< \Deltas$ so that the simple wave condition has been broken due to
linear degeneracy before singularity formation when $\Delta =
\Deltas$.  Figure \ref{fig:delta_plot} left, shows $1^+$-DSWs from
numerical simulations for $p = 2/3$ and varying $\Delta$.  A
transition to cavitation is observed for sufficiently large jumps as
predicted.  However, in the linearly degenerate regime $\Delta >
  \Delta_{\mathrm{l}}$, the simulation shows no noticeable change in
  the computed DSW structure.

\begin{figure}
  \centering
  \includegraphics{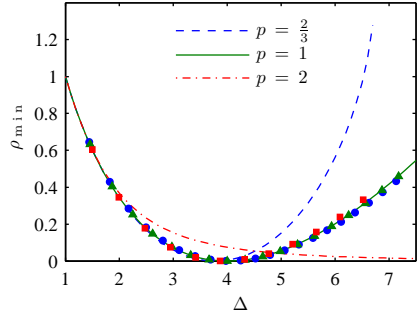}
  \caption{Power law gNLS DSW trailing soliton minimum $\rhomin$ as a
    function of the DSW jump $\Delta$ and nonlinearity exponent $p$.
    Theory (curves) and numerics ({\color{red} \fullsquare},
    {\color{grn} \fulltriangle}, {\color{blue} \fullcircle})
    significantly differ for $\Delta > \Deltav$ and $p \ne 1$.}
  \label{fig:soli_min_power_law}
\end{figure}

The soliton minimum $\rhomin$, determined by solving (\ref{eq:187})
with a numerical rootfinder, is shown in figure
\ref{fig:soli_min_power_law} for different choices of the nonlinearity
$p$ and jump heights (curves).  The simple DSW construction predicts
no cavitation for $p > 2$ and a sharply increasing soliton minimum as
the jump height is increased beyond $\Deltav(2/3)$.  Neither of these
behaviors are observed numerically.  Sufficiently large jumps in the
$p = 2$ case lead to cavitation. The numerically computed density
minima of the trailing edge soliton ({\color{red} \fullsquare},
{\color{grn} \fulltriangle}, {\color{blue} \fullcircle}), curiously,
lie on the integrable NLS curve.  It is not clear as to why this is
the case.

\subsection{Nonpolynomial Nonlinearity}
\label{sec:nonp-nonl}

Another particular form of the nonlinearity is now considered, that of
the nonpolynomial nonlinearity (\ref{eq:165}).  When inserted into
(\ref{eq:170}), this nonlinearity results in the non-separable ODE
parametrized by $\gamma$
\begin{equation}
  \label{eq:178}
  \frac{\rmd\alpha}{\rmd \rhobar} = -\frac{(1 + \alpha) [2 + 3 \gamma
    \rhobar + 2\alpha(2 + \gamma \rhobar)]}{4 \rhobar (1 + \gamma
    \rhobar) (1 + 2 \alpha)} .
\end{equation}
It is necessary to numerically solve (\ref{eq:178}) with the initial
conditions (\ref{eq:163}), or (\ref{eq:171}) with $\alpha \to
\alphat$, in order to recover DSW properties.  Since
\begin{equation*}
  \tilde{\alpha}_\mathrm{min}(\Delta) = -\frac{3}{2} + \frac{2}{2 + \gamma
    \Delta} < - \frac{1}{2}, \qquad \gamma \Delta > 0,
\end{equation*}
and $\alphat(\Delta)$ is a decreasing function for $\alphat >
\alphat_{\mathrm{min}}$, singularity formation ($\alphat \to -1/2$) is
guaranteed for $\gamma > 0$ and $\Delta$ sufficiently large.  Linear
degeneracy, however, occurs before this when
\begin{equation*}
  \alphat(\Delta_\mathrm{l}) = - \frac{2 + \gamma
    \Delta_{\mathrm{l}}}{4 + 4 \gamma \Delta_{\mathrm{l}}} >
  -\frac{1}{2} .
\end{equation*}
Note that for $\gamma \gg 1$, the characteristic equation
(\ref{eq:178}) asymptotes to the power law nonlinearity equation
(\ref{eq:212}) with $p = 1/2$ so that the behavior in the large
$\gamma$ limit can be recovered directly from (\ref{eq:225}) and
(\ref{eq:226}).

\begin{figure}
  \centering
  \includegraphics{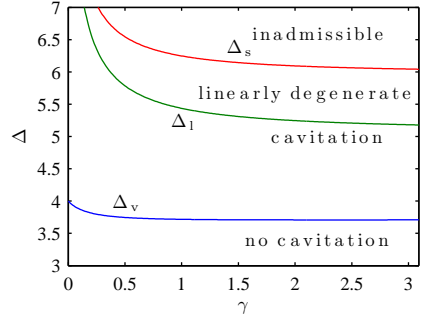}
  \caption{Singular $\Delta > \Deltas(\gamma)$, linearly degenerate
    $\Deltas(\gamma) > \Delta > \Delta_\mathrm{l}(\gamma)$ and
    cavitation $\Delta_\mathrm{l}(\gamma) > \Delta > \Deltav(\gamma)$
    regions for a simple wave led $1^+$-DSW with nonpolynomial
    nonlinearity.}
  \label{fig:delta_plot_np}
\end{figure}
Similar to the power law nonlinearity, the three distinguished values
for the shock jump $\Deltav < \Delta_\mathrm{l} < \Deltas$ divide
$(\gamma,\Delta)$ parameter space into various regions as shown in
figure \ref{fig:delta_plot_np}.  The large $\gamma$ behavior of the
distinguished values are, from (\ref{eq:189}), (\ref{eq:221}), and
(\ref{eq:183}) with $p = 1/2$
\begin{equation*}
  \fl
  \Deltav(\gamma) \sim 3.7, \qquad \Delta_\mathrm{l}(\gamma) \sim 5.1,
  \qquad \Deltas(\gamma) \sim 6.0, \qquad \gamma \gg 1 .
\end{equation*}
Cavitation and linear degeneracy are predicted to occur for $\gamma >
0$ with sufficiently large $\Delta$.  Additionally, the small $\gamma$
asymptotics can be determined as a regular perturbation to the NLS
solution (\ref{eq:217}).  The result is complex so it is not reported
here, deferring rather to the numerical solution of the ODE
(\ref{eq:178}).

\subsubsection{Numerical Results}
\label{sec:numerical-results-1}

\begin{figure}
  \centering
  \includegraphics{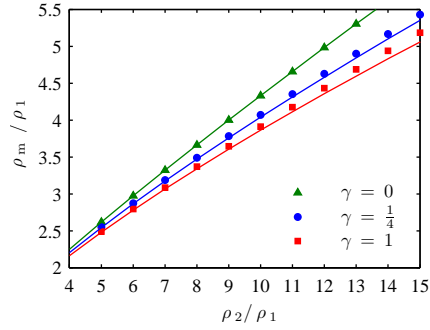}
  \caption{Predicted (\full) and computed ({\color{red} \fullsquare},
    {\color{grn} \fulltriangle}, {\color{blue} \fullcircle})
    intermediate states $\rhom$ resulting from the nonpolynomial gNLS
    shock tube problem.}
  \label{fig:shock_tube_intermediate_np}
\end{figure}

The shock tube problem for $\gamma \in \{0,1/4,1\}$ was solved
numerically.  The resulting DSW shock structure exhibits a
qualitatively similar pattern to the one shown in figure
\ref{fig:shock_tube_density}.  The computed intermediate density
$\rhom$ as compared with that resulting from the 1-DSW locus
(\ref{eq:295}) is shown in figure
\ref{fig:shock_tube_intermediate_np}.  As before, when deviating from
the integrable case with $\gamma > 0$, the 1-DSW locus provides a good
prediction to the intermediate density for weak to moderate jumps but
deviates in the large jump regime.

\begin{figure*}[!htb]
  \centering
  \subfigure[{Soliton edge speeds}\label{fig:dsw_soliton_speeds_np}]
  {\includegraphics{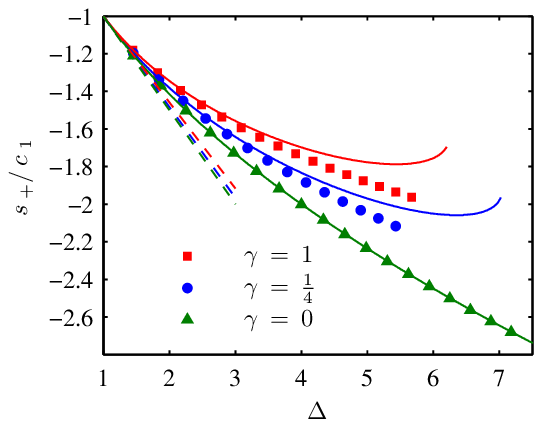}}
  \quad 
  \subfigure[{Linear wave edge speeds}\label{fig:dsw_linear_speeds_np}]
  {\includegraphics[scale=1]{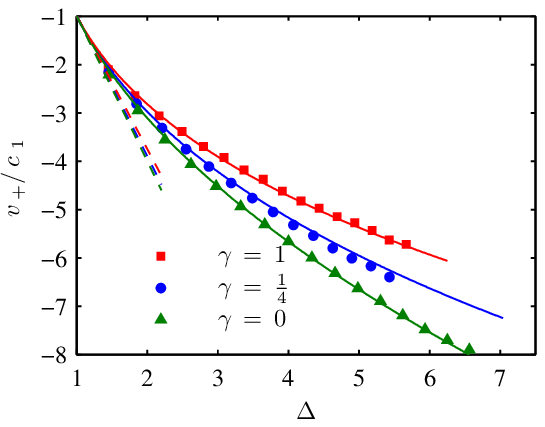}}
  \caption{Nonpolynomial gNLS 1-DSW speeds for varying density jump
    $\Delta$ and parameter $\gamma$.  Solid curves are from
    (\ref{eq:218}), (\ref{eq:213}), ({\color{red} \fullsquare},
    {\color{grn} \fulltriangle}, {\color{blue} \fullcircle}) result
    from numerical simulation, and (\dashed) are the weak DSW results
    (\ref{eq:124}), (\ref{eq:134}).} \label{fig:dsw_speeds_np}
\end{figure*}

The computed DSW speeds are pictured in figure \ref{fig:dsw_speeds_np}
for varying $\gamma$ and $\Delta$.  The weak DSW results (\dashed)
rapidly deviate from the predicted (\full) and computed ({\color{red}
  \fullsquare}, {\color{grn} \fulltriangle}, {\color{blue}
  \fullcircle}) results for $\Delta \gtrsim 1.5$.  The non-integrable
DSW speeds with finite $\gamma$ deviate from the DSW regularization in
the large jump regime $\Delta \gtrsim 3$, more so for the soliton edge
speeds.  The onset of linear degeneracy and singular behavior is
clear from the solid soliton speed curves, exhibiting minima at
$\Delta = \Delta_\mathrm{l}(\gamma)$ and an infinite derivative at
$\Delta = \Deltas(\gamma)$.

\begin{figure}
  \centering
  \includegraphics{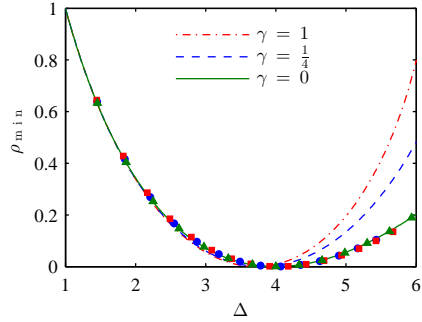}
  \caption{Soliton minimum $\rhomin$ as a function of the DSW jump and
    nonpolynomial nonlinearity coefficient $\gamma$.}
  \label{fig:soli_min_np}
\end{figure}

The soliton minimum $\rhomin$ is shown in figure \ref{fig:soli_min_np}.
As with the power law nonlinearity, the computed soliton trailing edge
minima lie approximately on the integrable NLS curve suggesting that
this behavior is not coincidental to a particular type of nonlinearity
chosen.

\subsection{Photorefractive Media}
\label{sec:phot-media-1}

\begin{figure}
  \centering
  \includegraphics[scale=1]{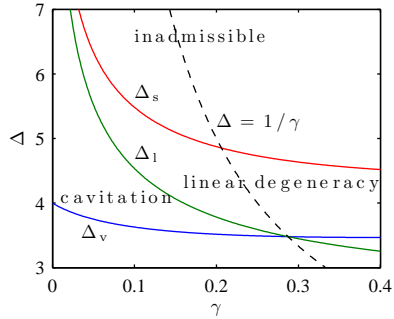}
  \caption{Validity of DSW modulation theory for photorefractive
    nonlinearity.  The dashed curve corresponds to $c'(\Delta) = 0$.}
  \label{fig:delta_plot_photo}
\end{figure}

The case of DSWs in photorefractive media (\ref{eq:84}) was studied in
detail in \cite{el_theory_2007}.  There, the modulation theory was
found to diverge for sufficiently large amplitude jumps and an
explanation in terms of the existence of cavitation was given.  Based
on the theory developed in section \ref{sec:breakd-simple-wave}, this
breakdown can be identified with linear degeneracy at the soliton
edge, hence the simple wave condition no longer holds.  A tell-tale
sign of breakdown is realized by the extremum of the shock speed with
respect to jump height in figure 7 of \cite{el_theory_2007}.  Figure
\ref{fig:delta_plot_photo} depicts the phase space $(\gamma,\Delta)$
divided into regions by the curves $\Delta_\mathrm{l}(\gamma)$,
$\Deltav(\gamma)$, and $\Deltas(\gamma)$.  In contrast to the previous
discussions, photorefractive nonlinearity only preserves the ordering
$\Deltav < \Delta_\mathrm{l} < \Deltas$ for $\gamma \lessapprox
0.287$.  However, the generic nonlinearity assumptions (\ref{eq:139})
corresponding to a convex pressure law or, equivalently, an increasing
sound speed are violated for jumps exceeding $\Delta > 1/\gamma$.  The
dashed curve $\Delta = 1/\gamma$ shown in figure
\ref{fig:delta_plot_photo} demonstrates that the change in ordering of
$\Delta_\mathrm{l}$ and $\Deltav$ occurs when $c'(\Delta) = 0$.

\section{Large Amplitude DSWs with Negative Dispersion}
\label{sec:large-amplitude-dsws}

Dispersive shock waves for fully nonlinear shallow water waves and
ion-acoustic plasma have been studied elsewhere
\cite{el_resolution_2005,el_unsteady_2006}.  It is worth commenting on
these results in light of the theory presented here.  Both systems
exhibit alternative negative dispersion regularizations of the Euler
equations in contrast to the positive dispersion of gNLS.

In shallow water with zero surface tension $\sigma = 0$
\cite{el_unsteady_2006}, the linear degeneracy condition
(\ref{eq:137}) occurs at the trailing edge for relatively small jumps
$\Delta = \Delta_\mathrm{l} \approx 1.43$.  For $\Delta >
\Delta_\mathrm{l}$, the simple wave theory breaks down and, indeed,
the numerically computed soliton edge speed begins to noticeably
deviate from its theoretical value (see figures 3 and 4 in
\cite{el_unsteady_2006}).  It is curious that the numerically computed
linear edge speed remains fairly accurate.  For $\sigma = 0$, zero
dispersion does not occur at either edge, nor does linear degeneracy
at the soliton edge.  For $\sigma > 0$, both $\omega_{kk} = 0$ and
$(\omegat/\kt)_{\kt} = 0$ when $\rho = \sqrt{3 \sigma}$, offering a
potential route to gradient catastrophe in the Whitham modulation
equations for appropriate DSW jumps.

Ion-acoustic plasma DSWs also exhibit linear degeneracy at the linear
wave edge for appropriate jump heights.  The linear edge speed plotted
in figure 5 of \cite{el_resolution_2005} for a $2^-$-DSW from $\rho_1
= \Delta$ to $\rho_2 = 1$ exhibits the requisite minimum as $\Delta$
is varied (recall (\ref{eq:107})).  A numerical computation shows that
the minimum occurs when $\Delta = \Delta_{\mathrm{l}} \approx 1.41$.
Zero dispersion at either edge and linear degeneracy at the soliton
edge do not occur by calculation of the criteria from section
\ref{sec:breakd-simple-wave}.

The particular dispersive Eulerian fluids discussed here demonstrate
the importance of the breakdown criteria (\ref{eq:106}),
(\ref{eq:107}) in both negative and positive dispersion cases.

\section{Conclusion}
\label{sec:conclusion-1}

Shock waves in the dispersively regularized isentropic Euler
$P$-system under modest assumptions were constructed using the
Whitham-El simple wave closure technique.  A complete, explicit
characterization of admissible, weak DSWs was shown to depend only on
the pressure law and convexity or concavity of the dispersion
relation.  Linear degeneracy of the modulation equations and zero
dispersion leading to gradient catastrophe in the modulation equations
were identified as causes of the breakdown of the simple wave
assumption.  Simple tests in terms of extrema of the DSW leading or
trailing edge speeds for these behaviors were elucidated.  Large
amplitude DSWs were constructed for the case of the gNLS equation
modeling super and optical fluids.  Comparisons with careful numerical
simulations of the shock tube problem reveal excellent agreement with
theory in the weak to moderate jump regime.  Deviation occurred in the
large jump regime with linear degeneracy in the modulation equations
at the DSW soliton edge a proximate cause for certain pressure laws.

While there are a number of parallels between classical, viscous
Eulerian fluids and dispersive Eulerian fluids, this work has
demonstrated that DSWs exhibit distinct physical and mathematical
behavior.  Physically, the generation of oscillations leads to an
expanding oscillatory region with two speeds in contrast to localized,
classical shock fronts propagating as traveling waves.  Positive
dispersion fluids can exhibit DSWs with cavitation while negative
dispersion fluids admit DSWs with backflow.  Mathematically, weak
Eulerian DSWs and shocks exhibit universal behavior, depending only
upon the sign of dispersion and the pressure law.  In the large
amplitude regime, universality is maintained for classical viscous
shock jump conditions, which are the same for a large class of
dissipative regularizations.  However, due to their nonlocal nature
and the weak limit involved, large amplitude DSWs crucially depend
upon the particular form of the dispersion.  Furthermore,
admissibility of simple wave led DSWs is much more subtle than the
elegantly stated Lax entropy conditions for classical shocks.

\ack

The author gratefully acknowledges financial support from the National
Science Foundation via DMS-1008973.

\appendix
\section*{Appendix:  Numerical Methods}
\setcounter{section}{1}

The numerical solution of the gNLS equation (\ref{eq:83}) for the
shock tube problem, the initial step in density
\begin{equation}
  \label{eq:104}
  \psi(x,0) = \left \{
    \begin{array}{ll}
      1 & x < x_0 \\
      \sqrt{\rho_2} & x > x_0
    \end{array} \right . , \qquad \rho_2 > 1,
\end{equation}
is briefly described here.  A pseudospectral, time-splitting method is
implemented for the accurate solution of long time evolution for $x\in
(0,L)$.  The initial data (\ref{eq:104}) is smoothed by use of the
hyperbolic tangent initial condition
\begin{equation*}
  \psi(x,0) = \left \{ \frac{1}{2} [1 + \tanh(x_0-x)](1 - \rho_2) +
    \rho_2 \right \}^{1/2} ,
\end{equation*}
where $x_0 = L/2$.  Time stepping proceeds by use of second order
Strang splitting \cite{strang_construction_1968} where the linear PDE
\begin{equation}
  \label{eq:108}
  \rmi \frac{\partial \psi_\mathrm{L}}{\partial t} = -\frac{1}{2}
  \frac{\partial^2 \psi_\mathrm{L}}{\partial x^2}, \qquad
  \psi_\mathrm{L}(x,t) = \psi(x,t),
\end{equation}
is advanced half a time step $\Delta t/2$ exactly followed by a full
time step of the nonlinear ODE
\begin{equation}
  \label{eq:109}
  \rmi \frac{\partial \psi_{\mathrm{NL}}}{\partial t} =
  f(|\psi_{\mathrm{NL}}|^2) \psi_{\mathrm{NL}}, \qquad
  \psi_{\mathrm{NL}}(x,t) = \psi_\mathrm{L}(x,t+\Delta t/2) .
\end{equation}
The linear PDE is then advanced half a time step with the initial data
$\psi_\mathrm{L}(x,t+\Delta t/2) = \psi_{\mathrm{NL}}(x,t+\Delta t)$
giving the second order accurate approximation of $\psi(x,t+\Delta t)
\approx \psi_\mathrm{L}(x, \Delta t)$.  Equation (\ref{eq:108}) is
projected onto a truncated cosine basis of $N$ terms that maintains
Neumann ($\psi_x = 0$) boundary conditions, computed efficiently via
the FFT, and integrated explicitly in time.  The nonlinear ODE
(\ref{eq:109}) conserves $|\psi_{\mathrm{NL}}|^2$ so is also
integrated explicitly in time.  The parameter $\Delta x = L/N$ is the
spatial grid spacing of the grid points $x_j = \Delta x (j - 1/2)$, $j
= 1,2,\ldots,N$.  The accuracy of the solution is monitored by
computing the relative deviation in the conserved $L^2$ norm $E(t) =
\int_{\R} | \psi(x,t) |^2 \rmd x$, $E_{\mathrm{rel}} =
|E(t_\mathrm{f})-E(0)|/E(0)$ where $t_\mathrm{f}$ is the final time.
All computations presented here exhibit $E_{\mathrm{rel}} < 10^{-8}$.
Also, the accurate spatial resolution of the oscillatory structures is
supported by the fact that the coefficient of the largest wavenumber
in the cosine series is less than $5 \cdot 10^{-10}$ (often times much
less).  The numerical parameters $L$, $N$, $\Delta t$, and
$t_\mathrm{f}$ vary depending upon the nonlinearity strength and jump
height.  For example, for power law gNLS with $p = 2$ and $\rho_2 \ge
11$, $N = 2^{16}$, $L = 1200$, $\Delta t = 0.0002$, and $t_\mathrm{f}
= 30$.  Whereas, for $p = 2/3$ with $\rho_2 = 2$, $N = 3 \cdot
2^{14}$, $L = 3000$, $\Delta t = 0.002$, and $t_\mathrm{f} = 500$.

The extraction of the DSW speeds $v_+$, $s_+$, and minimum density
$\rhomin$ is performed as follows.  The precise location of the DSW
soliton trailing edge is computed by creating a local cubic spline
interpolant through the computed grid points in the neighborhood of
the dark soliton minimum.  A root finder is applied to the derivative
of this interpolant in order to extract the off-grid location of the
soliton edge $x_s(t)$ and $\rhomin \equiv
|\psi(x_s(t_\mathrm{f}),t_\mathrm{f})|^2$.  The slope of a linear
least square fit through $x_s(t_j)$ for $j = 1, \ldots, 100$
equispaced $t_j \in [t_\mathrm{f}-1,t_\mathrm{f}]$ determines $s_+$.
For the leading, linear wave edge, an envelope function is determined
by least squares fitting two lines, each through about 30 local maxima
and minima, respectively, of the DSW density in the vicinity of the
trailing edge.  The extrema are computed the same as for the soliton
minimum.  The point of intersection of these two lines is the location
of the linear wave edge $x_v(t)$.  The same fitting procedure as was
used to determine $s_+$ from $x_s(t)$ is used to extract $v_+$ from
$x_v(t)$.

\section*{References}


\end{document}